# Resolving the Structural Duality of Graphene Grain Boundaries

*Haojie Guo*, Emiliano Ventura-Macías, Mariano D. Jiménez-Sánchez, Nicoleta Nicoara, Pierre Mallet, Jean-Yves Veuillen, Vincent T. Renard, Antonio J. Martínez-Galera, Pablo Pou, Julio Gómez-Herrero, Rubén Pérez, Iván Brihuega**

E-mail: haojie.guo@dipc.org; ivan.brihuega@uam.es

*H. Guo, M. D. Jiménez-Sánchez, J. Gómez-Herrero, I. Brihuega*
Departamento de Física de la Materia Condensada, Universidad Autónoma de Madrid, Madrid, Spain
*E. Ventura-Macías, P. Pou, R. Pérez*
Departamento de Física Teórica de la Materia Condensada, Universidad Autónoma de Madrid, Madrid, Spain
*N. Nicoara*
International Iberian Nanotechnology Laboratory (INL), Braga, Portugal
*P. Mallet, J.-Y. Veuillen,*
Université Grenoble Alpes, Institut NEEL, Grenoble, France
Centre National de la Recherche Scientifique (CNRS), Institut NEEL, Grenoble, France
*V. T. Renard*
Univ. Grenoble Alpes, CEA, Grenoble INP, IRIG, PHELIQS, 38000 Grenoble, France.
*A. J. Martínez-Galera*
Departamento de Física de Materiales, Universidad Autónoma de Madrid, Madrid, Spain
*A. J. Martínez-Galera, J. Gómez-Herrero, I. Brihuega*
Instituto Nicolás Cabrera (INC), Universidad Autónoma de Madrid, Madrid, Spain
*A. J. Martínez-Galera, P. Pou, J. Gómez-Herrero, R. Pérez, I. Brihuega*
Condensed Matter Physics Center (IFIMAC), Madrid, Spain
*H. Guo*
Present Address: Donostia International Physics Center (DIPC), San Sebastián, Spain

**Abstract:** Grain boundaries (GBs) are ubiquitous in large-scale graphene samples, playing a crucial role in their overall performance. Due to their complexity, they are usually investigated as model structures, under the assumption of a fully relaxed interface. Here, we present cantilever-based non-contact atomic force microscopy (ncAFM) as a suitable technique to resolve, atom by atom, the complete structure of these linear defects. Our experimental findings reveal a richer scenario than expected, with the coexistence of energetically stable and metastable graphene GBs. Although both GBs are structurally composed of pentagonal and heptagonal like rings, they can be differentiated by the irregular geometric shapes present in the metastable boundaries. Theoretical modeling and simulated ncAFM images, accounting for the experimental data, show that metastable GBs form under compressive uniaxial strain and exhibit vertical corrugation, whereas stable GBs remain in a fully relaxed, flat configuration. By locally introducing energy with the AFM tip, we show the possibility to manipulate the metastable GBs, driving them toward their minimum energy configuration. Notably, our high-resolution ncAFM images reveal a clear dichotomy: while the structural distortions of metastable grain boundaries are confined to just a few atoms, their impact on graphene's properties extends over significantly larger length scales.

# 1. Introduction

Mechanically exfoliated graphene is probably the most perfect material ever created, with a nearly defect-free crystalline structure. This exceptional structural quality translates into outstanding electronic, mechanical and optical properties and an unprecedented chemical inertness.[1-7] However, such crystalline perfection is not achievable at industrial scales, where large-scale production methods inevitably introduce defects during graphene growth.[8-15] The mass production of graphene typically relies on chemical synthesis techniques, such as chemical vapor deposition (CVD), epitaxial growth on silicon carbide (SiC), and the chemical reduction of graphene oxide.[16-23] These methods yield polycrystalline graphene in which grain boundaries (GBs) disrupt the lattice structure, playing a pivotal role in the electrical, mechanical, thermal, chemical, and magnetic characteristics of the material.[8-10,17,24-41] Consequently, thorough characterization and control of these linear defects are not only essential for understanding the true properties of large-scale produced graphene, but they also represent a remarkable opportunity to tailor and enhance its functionalities. This approach extends beyond graphene, offering similar opportunities for other non-carbon 2D materials.

While most studies on the impact of GBs on graphene properties rely on idealized grain boundary models, consisting of alternating fully planar pentagon-heptagon (5-7) ring pairs in large misaligned grain angles,[10,25,42-44] considerable efforts have been made in predicting realistic GB structures. These efforts underscore the complexity of the task, influenced by factors such as growth conditions, substrate interactions, and local stresses. In this sense, advances in experimental techniques, like transmission electron microscopy (TEM) and scanning tunneling microscopy (STM), have been crucial for accessing the atomic-scale structure of 2D materials and characterizing their defects.[18,24,26,40,45-52] However, these methods are not without limitations. TEM typically requires freestanding and electron transparent samples, and the electron beam can induce structural transitions or damage, particularly in systems with delicate energy balances.[26,52-54] In STM, the tunneling current reflects a convolution of topographic features and the local density of states. For atomic-scale defects in graphene, this poses a challenge, as ($\sqrt{3}\times\sqrt{3}$)-R30° (R3) patterns arising from intervalley scattering processes obscure the resolution of the actual atomic structure.[18,22,40,55]

Since the pioneering work by Gross *et al.*,[56] the use of ultra-high vacuum (UHV) non-contact atomic force microscopy (ncAFM) with small oscillation amplitudes and tips functionalized with CO or Xe has enabled the unambiguous resolution of the chemical structure of various molecules and graphene nanoribbons.[56-61] However, these studies typically required systems adsorbed on metallic substrates or ultrathin insulating layers adjacent to bare metallic regions, which serve as reservoirs for individual CO or Xe molecules needed to functionalize the AFM tip.[56-62] On chemically inert substrates as graphene, direct *in-situ* functionalization is not feasible as the low reactivity of 2D material prevents the adsorption of such functionalizing entities. The alternative requires a procedure consisting of first performing *ex-situ* tip functionalization with CO or Xe on metallic substrate and then switching back to the target graphene substrate[63-65].

In this work, we extend the above methodology to significantly larger modulation amplitudes (20-30 nm) ncAFM and using a tip functionalization process that results in an O-terminated tip-apex directly on graphene substrates (see Methods section). We demonstrate that cantilever-based ncAFM can resolve atom by atom, the real structure of defects in graphene layers grown on SiC(000-1). Focusing on naturally existing GBs, our technique reveals with unprecedented resolution the intrinsic atomic structure of two types of graphene GBs on the surfaces: metastable and stable (ideal) ones. By comparing our experimental images with theoretical modeling and simulated ncAFM images, we ascribe the formation of metastable GBs by the presence of compressive strain that induces vertical corrugation. In contrast, stable GBs maintain a completely planar and relaxed structural configuration. We experimentally find that the impact of metastable grain boundaries is not confined to the boundary region itself but also alters the graphene structure over many nanometers. Additionally, we demonstrate that localized mechanical energy injection via the cantilever tip can effectively manipulate metastable GBs, driving them into a stable, flat, planar configuration.

## 2. Results

We first explored the use of cantilever-based ncAFM for high-resolution imaging of GBs on graphene surfaces (see Methods for details). **Figure 1**a shows a constant-height ncAFM image of the boundary between two rotational graphene domains (a full-page view is provided in Figure S1 in the SI). In addition to resolving the hexagonal lattices of both graphene domains, in the middle of the image the GB region reveals an alternating sequence of pentagonal and heptagonal carbon rings. For clarity, these non-hexagonal rings are highlighted with green (pentagons) and blue (heptagons) dots. The raw ncAFM images were processed to suppress low-frequency moiré pattern signals, arising from the rotational misalignment between two stacked graphene layers that results in a larger-scale interference superlattice, and enhancing thus the visibility of chemical bonds (see Figure S2 and Figure S3 in the SI).

The observed alternating 5-7 ring configuration aligns with theoretical predictions, which identify this motif as the lowest-energy structural arrangement for high-angle GBs (e.g., θ ~ 32° as studied here).[25,42-44] Figure 1b illustrates a schematic model corroborating this result. It shows how a 32.2° in-plane twists between two graphene sheets (lattice constant 2.46 Å) naturally generates the experimentally resolved 5-7 dislocation pairs. This 5-7 geometry agrees with prior studies based on aberration-corrected transmission electron microscopy (AC-TEM) for free standing graphene samples.[26,45-47,49,52] Notably, the enhanced structural clarity and resolution achieved in this work represents a significant advancement and marks a new milestone in high-resolution imaging of C-based nanostructures.

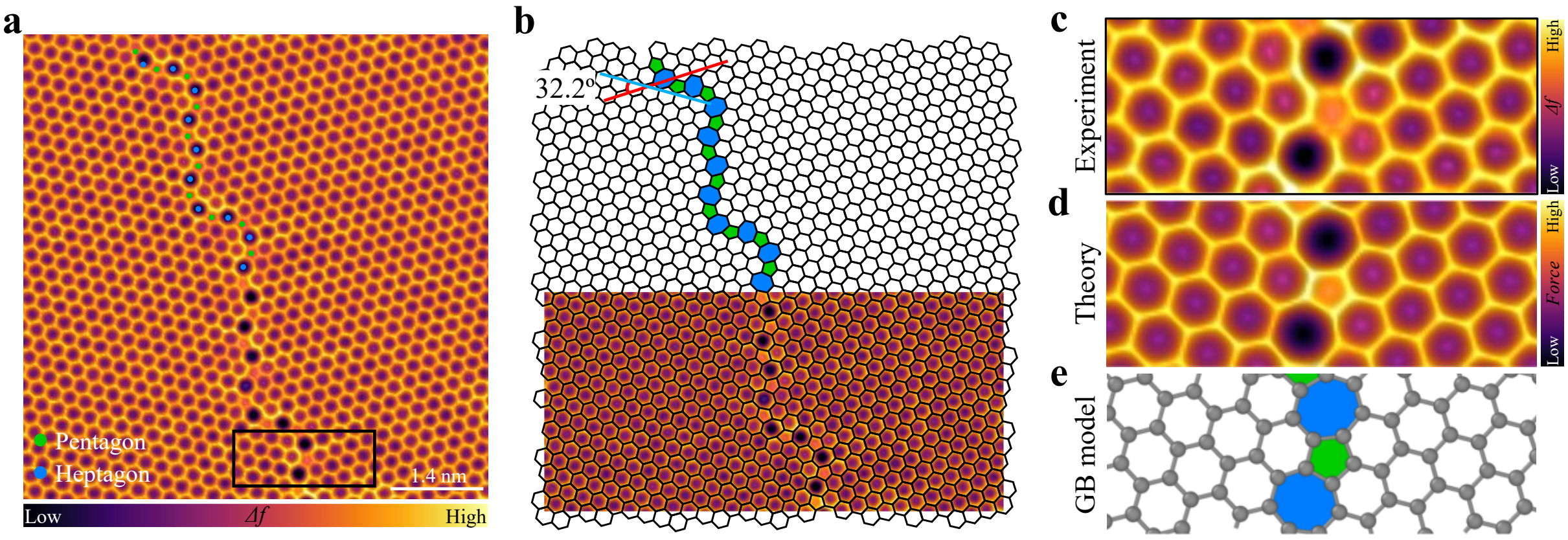

**Figure 1. High-resolution imaging of the atomic structure of graphene grain boundary.** a) Constant height ncAFM image showing a grain boundary region delimitating two graphene rotational domains, where the atomic structure of the GB can be inferred. FFT filtering process and 3D effect have been applied to the image. b) Schematic representation that reproduces the main characteristics of the GB shown in (a) by joining together two graphene domains with a misorientation angle of 32.2º. The experimental measured ncAFM image depicted in (a) is superimposed below the graphene lattice for comparison. c) Magnified view within the area indicated by the black rectangle in (a). The pentagonal and heptagonal rings are clearly observed. FFT filtering process and 3D effect have been applied to the image. d) Simulated constant height ncAFM image of a GB using a CO tip and taking into consideration the effect of tip relaxation ($k$ = 1.5 eV/rad$^2$). A 3D effect has been applied to the image. e) Optimized atomic structure of an ideal grain boundary, which separates two graphene domains with a misorientation angle of 32.2º, used to simulate the ncAFM image shown in (d).

To gain further insights from our experimental constant height ncAFM images, we conducted direct comparisons with density functional theory (DFT) calculations for the GB structure using VASP[66] and AFM image simulations using the Full Density Based Model (FDBM)[67] (see also Methods section). Figure 1c displays a zoomed in view of the area marked by the black rectangle in Figure 1a. The main features in this image are accurately reproduced by the simulated image in Figure 1d (see Figure S5a in the SI for simulations at different tip-sample distances). This simulation corresponds to a fully relaxed, flat GB, with a precise 32.2° misorientation angle (see the GB model in Figure 1e and Supplementary note 2 in the SI for details). The rather rigid CO probe ($k$ = 1.5 eV/rad$^2$)[58,68,69] (see also Methods section) used in the AFM simulations provides an accurate description of the O-terminated tip suggested by the experimental contrast.

The exceptional agreement between our experimental ncAFM images and DFT calculations unambiguously identifies this specific GB, intrinsic to epitaxial graphene/SiC, as an ideal, fully relaxed configuration. Moreover, this result provides a strong foundation for further exploring and characterizing other potential geometrical configurations of graphene GBs in our as-grown samples. In **Figure 2**a, a constant-height ncAFM image measured in the same sample, but in a different region, shows a GB separating two graphene domains rotated by θ ~ 31º (an enlarged view is provided in Figure S7a in the SI), a similar angle as in Figure 1. A zoomed-in view of the area indicated by the blue rectangle in Figure 2a is presented in Figure 2b. Unlike the canonical 5-7 ring arrangement in Figure 1, this boundary exhibits structural irregularity with barely visible C atoms at the interface. Atomistic simulations incorporating strain and out-of-plane corrugation revealed the origin of this anomaly (see Methods section). The optimal match, shown in Figure 2c (see Figure S5b in the SI for simulations at different tip–sample distances), emerged from a compressed GB model (32.2° misorientation) with an in-plane uniaxial strain ($s$ = 25 pm), concentrated within the boundary region, inducing a strongly localized out-of-plane deformation ($h$ = −10 pm) in the GB area. This out-of-plane deformation is clearly visualized in the lateral view in Figure 2d and discussed in Supplementary note 2 in the SI. This specific configuration, with a 10 pm out-of-plane corrugation under 25 pm of in-plane strain, was selected as it provided the best agreement with the experimental AFM images in terms of contrast and atomic-scale features (see Figure S7 in the Supporting Information). The proposed model is further supported by energetic considerations from our DFT calculations (see further below).

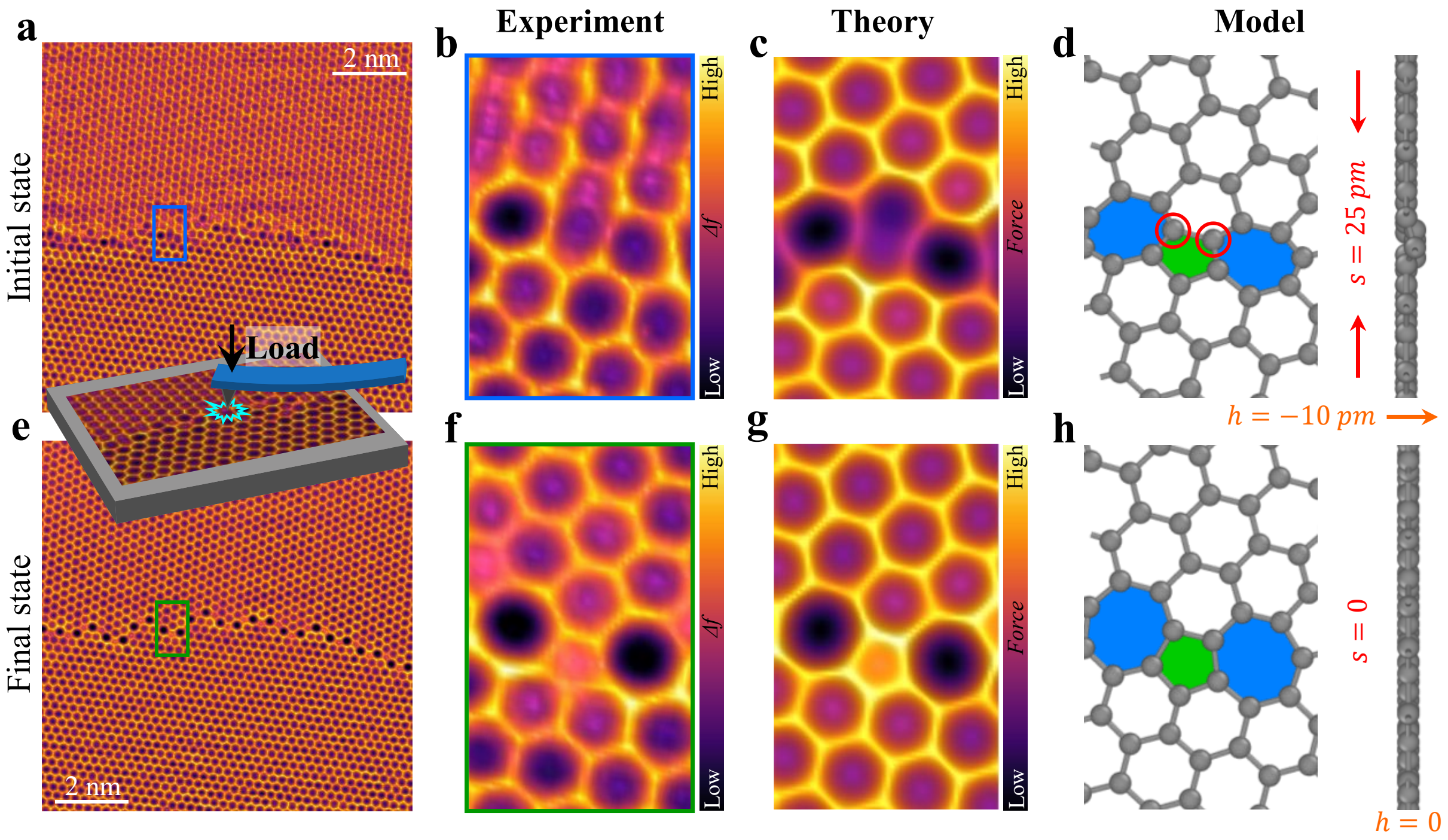

**Figure 2. Manipulation of metastable graphene grain boundary.** a,e) Corresponding constant height ncAFM images acquired in the same region before and after manipulating the GB with the ncAFM tip. Inset: 3D representation illustrating the process of grain boundary manipulation using the ncAFM tip, which consists of repeated indentation along different locations of the GB. b,f) Magnified view within the region delimited by the blue and green rectangle highlighted in (a,e), respectively. c,g) Theoretical simulated constant height ncAFM images that reproduce fairly well the feature of the grain boundary shown in the experimental images. A CO molecule ended tip-apex was employed in the simulation and the effect of tip relaxation ($k$ = 1.5 eV/rad$^2$) is considered. d,h) Grain boundary model used to simulate the theoretical ncAFM images. In (d), we considered a GB under the effect of a compressive strain $s$ = 25 pm, concentrated within the boundary region, and a vertical corrugation $h$ = -10 pm (red cycles indicate the two lowered atoms), while in (h), we considered an ideal fully planar GB. A FFT filtering process and 3D effect have been applied to all images, except for the theoretical images, which only have a 3D effect.

We extended our analysis to multiple graphene regions, conclusively demonstrating that as-grown graphene GBs on SiC coexist in both fully relaxed and strained configurations (see additional representative data in Figure S12 in the SI). These strained, corrugated configurations likely represent metastable GBs, where kinetic limitations during epitaxial growth hindered full structural relaxation. Crucially, this metastability suggests that targeted energy input—such as mechanical perturbation via the AFM tip—could drive these systems toward their thermodynamically favored states.

To evaluate the feasibility of this approach, we employed the AFM cantilever to locally inject mechanical energy into the metastable GBs, as the one shown in Figure 2a–d. We applied localized stress (indentation) by performing frequency shift *versus* distance ($\Delta f(z)$) curves at several different places along the GB area and its close vicinity (see the schematic in the inset of Figure 2 and Supplementary note 3 in the SI). Post-intervention ncAFM images (Figure 2e and enlarged view in Figure S7b in the SI) revealed a structural reorganization within the GB, with the previously irregular heptagons transitioning to a periodic 5-7 ring pattern (see the zoomed view in Figure 2f and Supplementary note 3 for a comparison of the initial and final GB states). The final reconfigured structure quantitatively matches the simulated ncAFM image of an ideal GB (Figure 2g), which was derived from a flat, strain-free model (Figure 2h; computational parameters are detailed in Supplementary note 2). Further example of tip-induced modification of a different metastable GB is provided in Figure S13 in the SI.

To experimentally quantify the atomic reorganization, we analyzed the structural differences between the initial and final GB configurations by overlaying ideal graphene lattices on both sides of the GB (**Figure 3**a–c). This comparison revealed that the as-grown GB exhibits an in-plane compression of about 20 pm localized just at the frontier between both grains, impeding the formation of regular flat 5-7 motifs. The complete strain localization at the GB area is highlighted by the empty dots in Figure 3b, which correspond to the C atoms that were displaced out-of-plane in the original metastable GB configuration and thus were barely visible in the ncAFM image shown in Figure 3a (see also Figure S14 in the SI). Notably, our ncAFM measurements are sensitive to these subtle variations, allowing us to distinguish differences in the local graphene structure across the different phases. These variations may originate from small charge density asymmetries, bond order differences, further emphasizing the power of ncAFM to probe not only topography but also the underlying electronic structure.[70]

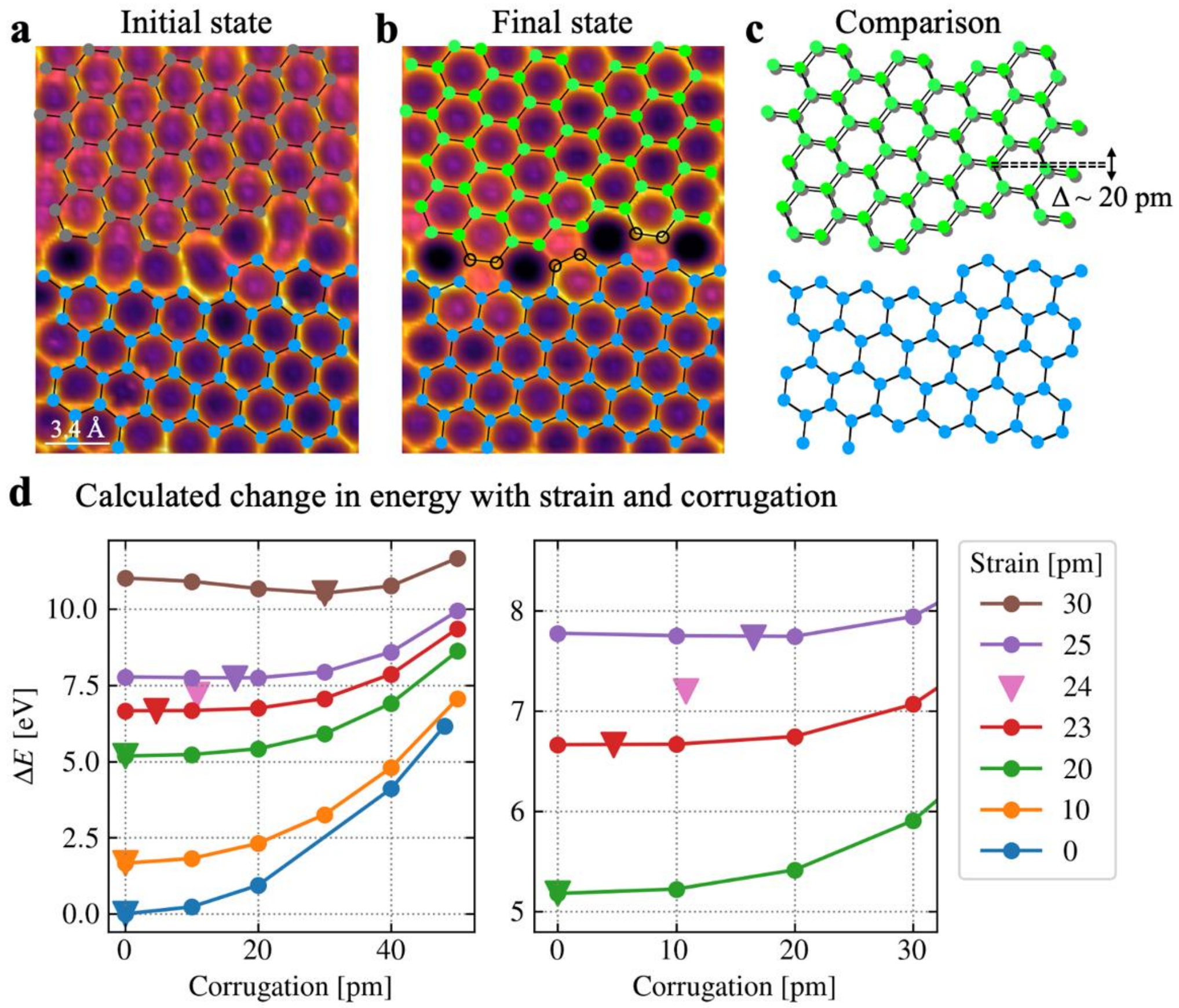

**Figure 3. Experimental evidence of strain on the grain boundary and the expected theoretical energetic cost.** a-b) Constant height ncAFM images corresponding to the GB discussed in Figure 2a,b, respectively. Two honeycomb lattices, marking the C atoms position, are superimposed on both graphene domains on each image. Those atoms that are corrugated initially are now visible after manipulation (see the empty circles). By comparing both graphene lattices before and after manipulation in (c), it can be inferred that there is a rigid shift (Δ ~ 20 pm) of the whole upper graphene lattice after manipulation. d) Calculated energy difference of the simulated GB systems (Figure S4 in the SI) under different corrugation ($x$-axis) and uniaxial strain (color) conditions (with no-strain and no-corrugation as reference). The strain was introduced by rigidly displacing one grain toward the other by a distance $s$, while simultaneously reducing the supercell dimension along the $x$-direction by the same amount. Each point is a single SCF calculation with the corresponding strain and corrugation. The inverted triangles represent the optimal corrugation (the one that minimizes the total energy) for a given strain. For a strain corresponding to $s$ = 24 pm only the minimized corrugation was calculated (~ 10 pm).

After the tip-induced manipulation, the compressive strain is released in form of a rigid shift of the graphene grain above the GB, which retracted by 20 pm, while the lower grain remained nearly stationary—a phenomenon we tentatively attribute to its anchoring by macroscale surface defects, such as wrinkles and beads observed in its vicinity (see Figure S15 in the SI). These defects likely act as pinning sites, kinetically trapping the GB in a metastable state during epitaxial growth. Overall, these findings demonstrate that the applied mechanical energy relieves the existing stress, in this case, by rigidly retracting the upper graphene grain enabling the metastable GB to revert to a fully flat, ideal and energetically favored configuration. Qualitatively, this idea agrees with our DFT calculations, which indicate that the strain-free ideal GB configuration has a lower energetic formation than the metastable one (Figure 3d). Moreover, as can be noted, our calculations show that a compressive strain of $s \sim 24$ pm yields a local energy minimum for a $h \sim -10$ pm corrugation, which further validates the proposed structural model of the metastable grain boundary discussed above.

While the exact energetic cost of converting metastable to stable GB configurations is difficult to determine (experimentally and theoretically), a qualitative estimate indicates that the nc-AFM cantilever can supply sufficient mechanical work to induce the transformation. Under our operating conditions ($k = 35$ N/m, $A_0$=20nm), the cantilever stores oscillation energy in the keV range; during controlled indentation, a fraction of this energy is transferred to the surface and partially dissipated via non-conservative processes (phonon excitation, plasticity, atomic rearrangements), see Figure S9 in the SI with the *F(z)* curves extracted from the characteristic *Δf(z)* data acquired during manipulation. These considerations support that localized tip-induced input is sufficient to overcome the metastability of as-grown GBs,

## 3. Discussion

Our theoretical simulations accurately reproduce our experimental ncAFM images of strained grain boundaries, enabling us, for example, to correlate the apparent absence of two C atoms at the edge of one grain with their downward displacement by approximately 10 pm (Supplementary note 2 in the SI). However, it is important to note that the idealized GB model used in our simulations captures only a localized structural unit and does not reflect the more complex, irregular character of the actual grain boundary observed experimentally. In the ncAFM images, the GB displays a non-periodic, zigzag-like morphology, with apparent directional segments typically spanning around three graphene unit cells. This zigzag-like morphology likely arises from local variations in strain and relaxation, which induce alternating configurations along the GB. In particular, we observe that for some segments, whose orientation is similar to the strain direction applied in our simulations, two edge atoms from one grain (e.g., the upper grain) appear to relax downward, while in adjacent segments the local strain orientation differs more substantially—deviating from the simulated strain profile—and the relaxation behavior becomes more complex or less well-defined. This alternation leads to an effective edge relaxation over extended GB segments, with one grain exhibiting a greater number of relaxed edge atoms. Thus, this relaxation profile acts analogously to an "effective edge", which has important implications for the electronic and atomic structure of the adjacent graphene grains. This is seen, for example, in Figures S14a and S16 in the SI, where the upper grain in the strained GB displays a peculiar bond texture in which one bond per carbon hexagon has higher intensity. On the contrary, this so-called quinoidal bond texture is not seen near the relaxed grain boundary. While the detailed description of this phenomenon is far beyond the scope of this study, we believe that an intuition of it may be found in the framework of Clar sextet theory.

Clar theory states that the π-electron of graphene is described by the superposition of three equivalent Clar formulas, in which every third carbon hexagon hosts a Clar sextet, preserving the overall aromatic character across the lattice.[71,72] The presence of defects or edges generally breaks this degeneracy, leading to a departure from the prototypical all-benzenoid configuration.[71-73] Interestingly, the grains boundary does not appear to break the aromaticity of adjacent graphene grains. Indeed, it is possible to construct Clar formulas preserving the threefold degeneracy of Clar phases on both sides of the grain boundary if the double bonds associated to its formation remain confined in it (Figure S17 in the SI). However, the aromaticity of the upper grain seems to be lifted in the presence of strain as evidenced by the quinoidal bond texture. We believe this can be understood as dramatic long range electronic reorganization due to a small change in the distribution of double bonds within the grain boundary as illustrated by Clar formulas consistent with the quinoidal bond texture in Figure S18. At the moment, we do not have sufficient understanding of the details of the grain boundary to explain the selection of this particular configuration over the other possible ones. We believe that specific theoretical treatment is needed to solve this issue in the future. Similarly, we acknowledge that our current work does not provide information on the electronic properties of the observed metastable GBs, which constitute an interesting avenue to be experimentally and theoretically explored in the future.

Overall, our study provides several significant novel aspects in the characterization of graphene surface and its GB. While previous works had determined the atomic structure of graphene GBs using TEM[26,45-47,49,52], this work constitutes the first to demonstrate that ncAFM can also resolve these structures. This alternative approach offers several advantages: it is suitable for semiconducting and insulating substrates, does not depend on material thickness, and can be applied to bulk material samples. Second, the study demonstrated for the first time that graphene GBs can also exist under metastable configurations, which is an aspect that has been largely overlooked until now, even theoretically. In addition, it is shown that the metastable grain boundaries can be driven to an energetically stable configuration through tip-induced manipulation. This result adds a new paradigm to the understanding of the structural properties of graphene GBs.

On the technical side, we have introduced a tip functionalization method that can be performed *in-situ* directly on the graphene substrates to achieve high-resolution atomic contrast imaging of its surface. While offering similar resolution output, it substantially simplifies the experimental procedure necessary in traditional *ex-situ* techniques that require first functionalizing the tip with CO or Xe on metallic surfaces and then switching back to the target graphene sample[63-65]. Given the success of O-terminated tips in imaging polar materials like oxides (Ruben: add references from Monig and Diebold), we expect this tip conditioning could be used for high-resolution imaging of other 2D materials like the Transition Metal Dichalcogenides (TMDs).

Finally, although earlier studies using ncAFM had resolved the atomic structure of graphene surfaces[64], none had focused specifically on grain boundaries. Moreover, these early results were obtained using tuning-fork based ncAFM systems with a small modulation amplitude (< 1 Å). This is in stark contrast to the cantilever-based ncAFM system set up with a large modulation amplitude (> 20 nm) employed in this work. Therefore, this work broadens the applicability and versatility of ncAFM.

## 4. Conclusion

In summary, we employed cantilever-based ncAFM under UHV at 5 K to characterize naturally occurring grain boundaries on polycrystalline graphene. High-resolution constant-height imaging directly revealed two atomic-scale GB configurations: metastable boundaries, identified by irregular pentagon–heptagon ring patterns, and ideal, fully planar relaxed boundaries. Supported by theoretical modeling and simulated ncAFM images, we showed that metastable GBs arise from local uniaxial compressive strain and vertical corrugation, while ideal GBs remain flat. By injecting localized mechanical energy with the ncAFM tip, we converted strained, metastable GBs into their stable, relaxed form. Together, these results establish ncAFM as a noninvasive, atomically precise technique for both resolving and manipulating defects in 2D materials. Importantly, metastable GBs not only introduce local scattering centers affecting charge/thermal transport or inducing local magnetic moments but also exert influence well beyond their immediate vicinity. Such long-range effects necessitate their inclusion in realistic polycrystalline graphene models, particularly for GB-containing devices.

## 5. Methods

*Sample preparation*: The graphene samples were grown under UHV condition by graphitization of the SiC(000-1) surfaces (Carbon face termination) following similar procedure described elsewhere.[18,22] The resulting samples are characterized by a multilayer graphene scheme, with typical thickness of ~ 3-5 layers, that features a highly disordered rotational stacking configuration between the successive graphene layers. This leads to an electronic decoupling of the π-bands between graphene sheets. Therefore, the first surface graphene layer can be considered as ideal graphene, with the Dirac point being close to the Fermi level (neutral doping).[74-76] Another characteristic of the samples used in the present work is that graphene surfaces present different crystallographic orientation. This triggers the natural formation of grain boundaries (GBs) at the frontiers that separate two graphene rotational domains (polycrystalline surface). Such regions are identified in this work by acquiring atomically resolved ncAFM images. Additionally, as a consequence of the rotational disorder of graphene layers grown on SiC(000-1),[18,22] moiré superstructures can be commonly found on the surface of Gr/SiC(000-1) samples (Figure S19 in the SI).

*ncAFM measurements*: Experiments were conducted at sample temperature of 5 K in an UHV system (base pressure ~ $10^{-10}$ Torr) that features two interconnected chambers. One of them is used for sample cleaning and preparation, while the other host a custom-built cantilever-based combined low temperature ncAFM/STM microscope.[77] An optical interferometer setup is employed to detect the dynamic of the cantilever.[78] AFM measurements were performed in the frequency modulation mode, while keeping a constant oscillation amplitude. Commercially available platinum coated Si cantilevers (PPP-NCLPt, Nanosensors™) were used as probes for the experiments. In this work, typical values of the cantilever's dynamic are: oscillation amplitude $A \sim 20$ nm, natural resonance frequency $f_0 \sim 1.7\times10^5$ Hz, quality factor $Q = \sim 6.2\times10^5$, cantilever stiffness $k = 35$ N/m. All ncAFM images were measured either in the constant height mode or in the constant frequency shift ($\Delta f$) mode, as it explicitly indicated on each figure. During data acquisition, a bias voltage ($V_{CPD}$) was applied to the sample to minimize the long-range electrostatic forces. STM measurements were performed under the same ncAFM experimental setup (the cantilever oscillation is kept active) but switching the feedback control channel now to the tunneling current. The topography STM images were measured in the constant current mode. All ncAFM/STM images were acquired and processed with the help of the WSxM software.[79]

*Tip-apex preparation*: Before performing high-resolution ncAFM imaging of the chemical structure of the GBs, the tip-apex of the cantilevers are prepared on several contaminants located on specific regions of our graphene samples (see Supplementary note 9 in the SI for more details). This is needed to achieve the ultimate atomic-scale resolution shown here, which is not accessible using the bare tip-apex of a commercial cantilever. Our procedure differs from the standard experimental techniques reported in the literature for the acquisition of bond-resolved images using ncAFM, which are based mainly on the controlled tip functionalization with CO molecules, and also with other inert terminations like a $CuO_x$ nanocluster or a Xe atom.[56,57,80] The contrast observed in the experiments strongly suggests an O-terminated apex. This identification is further supported by our simulations with a rigid CO probe (see Theoretical modelling section below), that provide a striking match with the experimental results. In the present case, we tentatively attribute the presence of oxygen at the apex to some kind of metallic oxides ($MO_x$) that we intentionally pick up during scanning from the contaminants present in our samples, although the ultimate chemical composition remains unknown. Nevertheless, we noted that such uncontrolled tip-apex functionalization can be performed on a daily basis, as it can be inferred from the multiple high-resolution images of different grain boundaries that we were able to measure and shown in this work. However, we emphasize that there could be some experimental challenges and constraints when these unknown contaminants are not present in other samples in the future.

Complementary, we have also shown that high atomic resolution of the grain boundaries obtained in the constant height ncAFM imaging mode is not attainable in the topographic constant $\Delta f$ acquisition mode (see Figure S21 in the SI). Similarly, we corroborate that it is not feasible to image the chemical structure of the graphene grain boundaries studied in this work with STM (see Figure S22 in the SI), which is in good agreement with previous works.[18,24,40,48,50,51]

*Theoretical modelling*: Simulations were carried out to determine the possible atomic structure of the grain boundary. These included DFT simulations to establish the energy landscape of strain and corrugation on the GB and AFM simulations to compare with observed experimental results. The structure for the 32.2º GB was adapted from reference[25]. The supercell (see Figure S4 in the SI) included two anti-parallel and identical grain boundaries along the y-axis with equidistance to their periodic images.

Experimental analysis (see Figure 3c) reveals that the initial state of the boundary appears locally compressed in the direction perpendicular to the GB. This strain is confined to the boundary region, which becomes narrower compared to the final unstrained configuration. The complexity of the system and the difficulty in determining realistic constraints, including the effects of underlying layers, the macroscopic strain distribution, etc., prevents the use of standard DFT calculations allowing full atomic relaxation. Instead, we constructed a simplified model that captures the essential features of the system. Starting from the ideal GB geometry, uniaxial strain was introduced by rigidly displacing one grain toward the other by a distance $s$, while simultaneously reducing the supercell dimension along the x-direction by the same amount. This approach concentrates the imposed strain within the boundary region, consistent with experimental observations. Keeping this strained configuration fixed, a set of single-point DFT calculations was performed for different vertical corrugation values to determine the corresponding free energy landscape. Corrugation was modeled by lowering the two C atoms shared by the 5-6 pair flanked by two 7 rings. In this way, we obtained the DFT free energy for each combination of strain and corrugation.

Calculations were performed with VASP[66] under the PBE-GGA approach to the exchange-correlation potential[81] and using a PAW pseudopotential[82]. The plane waves basis set was cutoff at 400 eV. Dispersion interactions were included with the DFT-D3 correction.[83] The supercell was sampled in the reciprocal space with a Monkhorst-Pack 3x5x1 k-points grid.[84] The reference energy was from the optimized grain boundary (no-strain, no-corrugation).

To validate the most likely structures and the experimental observations, ncAFM images were simulated under various combinations of strain and out-of-plane corrugation with the Full Density Based Model (FDBM).[67,85] This model retains the DFT accuracy in the calculation of forces with inert molecular probes and provides the computational efficiency needed to simulate the entire ncAFM image. Application to different systems shows that FDBM captures the most important features in ncAFM images, providing an accurate description of the observed contrast. These features arise from the interplay of three different force contributions: short-range Pauli repulsion, electrostatics, and dispersion forces, together with probe deflection induced by the interaction.

Based on the observed resolution and contrast features, and the tip-apex preparation followed in the experiments, the most probable tip configuration corresponds to a metal oxide $MO_x$ apex terminated by a single oxygen atom. To approximate this configuration in simulations, a CO-functionalized probe was employed as implemented in our well-tested FDBM framework. This approach effectively captures the main characteristics expected from an oxygen-terminated $MO_x$ tip. In the case of metal tips functionalized with a CO molecule at the apex, the high lateral flexibility of the CO probe strongly contributes to the sharp bond contrast observed in ncAFM images at short distances.[58,60] To better match the experimental contrast observed with the more rigid O-terminated $MO_x$ tip, we employed a stiffer torsional spring constant ($k$) of 1.5 eV/rad$^2$, in contrast to the standard value of 0.4 eV/rad$^2$ typically used for CO-metal probes.[67,85] The short-range Pauli interaction in FDBM has been calculated with parameters $\alpha = 1.06$ and $V = 35.2$ eV Å$^{3(2\alpha-1)}$.

## Acknowledgements

I.B. acknowledges financial support from Spanish Ministry of Science and Innovation (MICINN) through grant No. PID2023-149106NB-I00, the Comunidad de Madrid and the Spanish State through the Recovery, Transformation and Resilience Plan [Materiales Disruptivos Bidimensionales (2D), (MAD2DCM)-UAM Materiales Avanzados] and the NMAT2D-CM program under grant S2018/NMT-4511. J.G.H. acknowledges financial support from MICINN through grants No. PID2019-106268GB-C3 and PID2022-138908NB-C32. H.G. and J.G.H. acknowledge financial support from the Comunidad de Madrid and Universidad Autónoma de Madrid through grant No. PEJ-2020-AI/IND-17765. H.G. acknowledges financial support from MICINN through grant No. PID2020-116619GA-C22. A.J.M.G. acknowledges financial support from MICINN through grants No. PID2020-116619GA-C22, CNS2024-154772, TED2021-131788A-I0, through the Ramón y Cajal program under grant No. RYC2022-038197-I, and from the Comunidad de Madrid and the Universidad Autónoma de Madrid through grant No. SI3/PJI/2021-00500. E.V.M., P.P. and R.P. acknowledge financial support from MICINN through grants No. PID2020-115864RB-I00 and PID2023-149150OB-I00. E.V.M., P.P., J.G.H., R.P., and I.B. acknowledge support from the María de Maeztu Program for Units of Excellence in R&D (grant No. CEX2023–001316-M). The Spanish Supercomputing Network (RES) is acknowledged for providing computational resources at the Marenostrum Supercomputer (BSC, Barcelona).

**Data Availability Statement**

The data that support the findings of this study are available from the corresponding authors upon reasonable request.

**Competing interests**

The authors declare no competing interests.

**Supporting Information**

Supporting Information is available from the Wiley Online Library or from the author.

# ToC

Cantilever ncAFM resolves the atomic structure of grain boundaries in graphene:, revealing coexisting stable and metastable types. Both contain pentagon/heptagon defects, but metastable GBs show irregular geometries. Modeling shows metastable GBs form under compression, exhibiting vertical corrugation, while stable GBs are flat. Metastable GBs can be manipulated towards stability. Their localized distortions impact properties over extended scales.

**Resolving the structural duality of graphene grain boundaries**

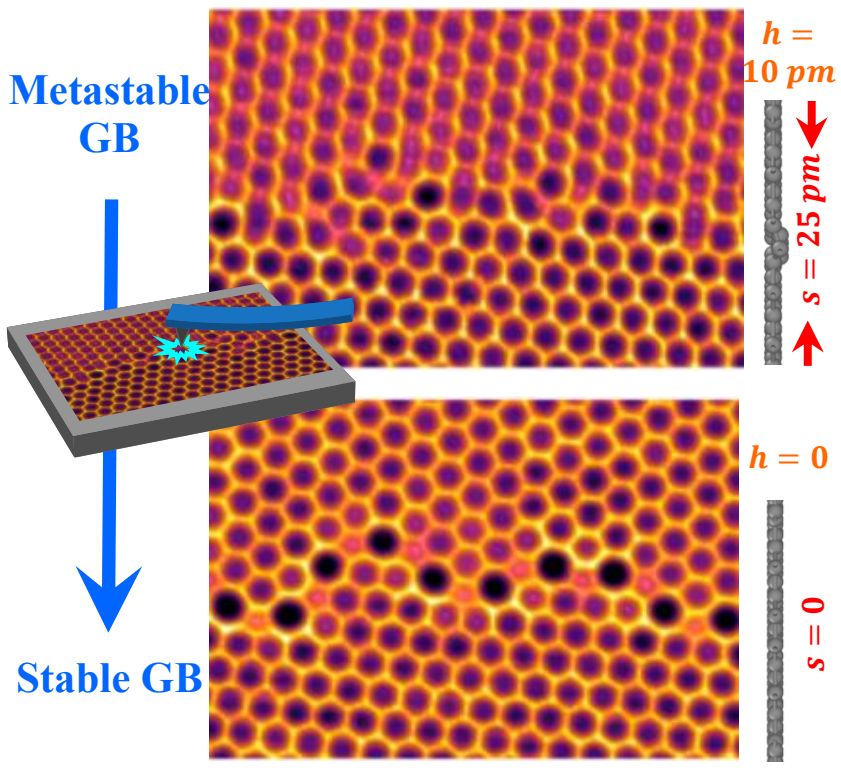

# Supporting Information for

## Resolving the Structural Duality of Graphene Grain Boundaries

*Haojie Guo*, Emiliano Ventura-Macías, Mariano D. Jiménez-Sánchez, Nicoleta Nicoara, Pierre Mallet, Jean-Yves Veuillen, Vincent T. Renard, Antonio J. Martínez-Galera, Pablo Pou, Julio Gómez-Herrero, Rubén Pérez, Iván Brihuega**

**Corresponding authors:* haojie.guo@dipc.org; ivan.brihuega@uam.es

**This PDF includes:**

**Supplementary note 1: Data processing of the experimental constant-height ncAFM images**

In this note, we disclose in detail the filtering procedure applied to some of the ncAFM images presented in the main manuscript (Figures 1–3) and in the supplementary information (e.g., Figure S1). This procedure was designed to enhance the visibility of features within the grain boundary regions, which are the central focus of this study. The approach involves a two-step method, as illustrated with the example shown in Figure S2. We begin with the experimentally acquired raw image (Figure S2a), which reveals two graphene rotational domains separated by a grain boundary region, along with superimposed moiré patterns. We first applied an FFT filtering to eliminate the low-frequency (long-wavelength in real space) signals associated with the moiré patterns (Figure S2b). Following this, a 3D illumination effect was employed to enhance contrast along the *z*-axis, thereby improving the visibility of the chemical bonds (Figure S2c). The resulting image corresponds to those shown in Figure 1 and Figure S1. By comparing the images shown in Figure S2, we observe that our data processing method minimally impacts the grain boundary (GB) region, which is the primary area of interest in this work.

As we stated above, the FFT filtering process removes only the low-frequency signals from the non-contact constant-height ncAFM images. These low-frequency components are typically linked to the moiré patterns observed on the graphene surface. Therefore, by applying FFT filtering, we effectively suppress only the moiré pattern signal contrast without altering the intrinsic features of the topmost graphene layer, as visually demonstrated in Figure S3. This approach ensures enhanced feature visibility while preserving the integrity of the GB regions.

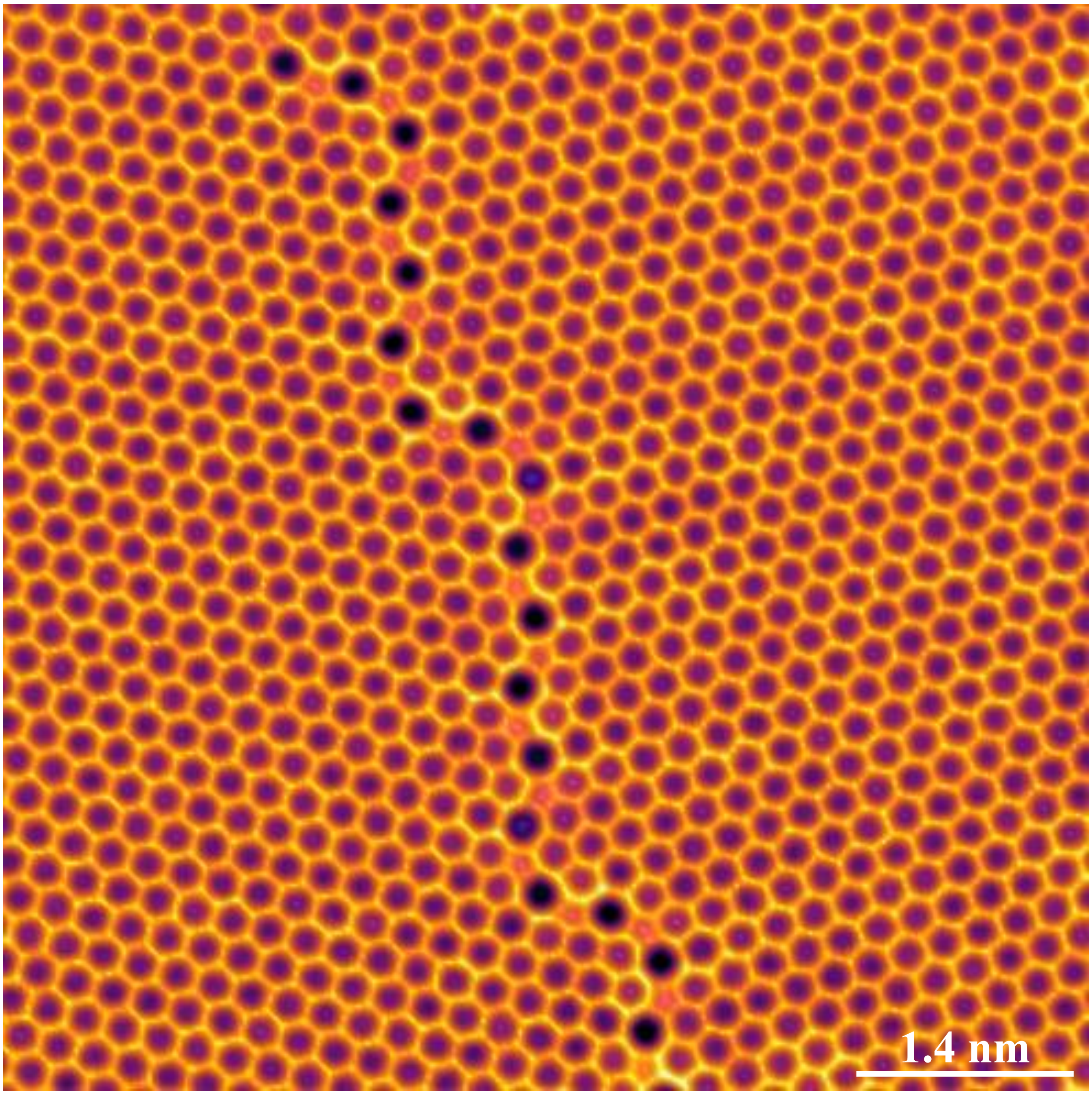

**Figure S1.** Enlarged view of the same constant height ncAFM image shown in Figure 1a.

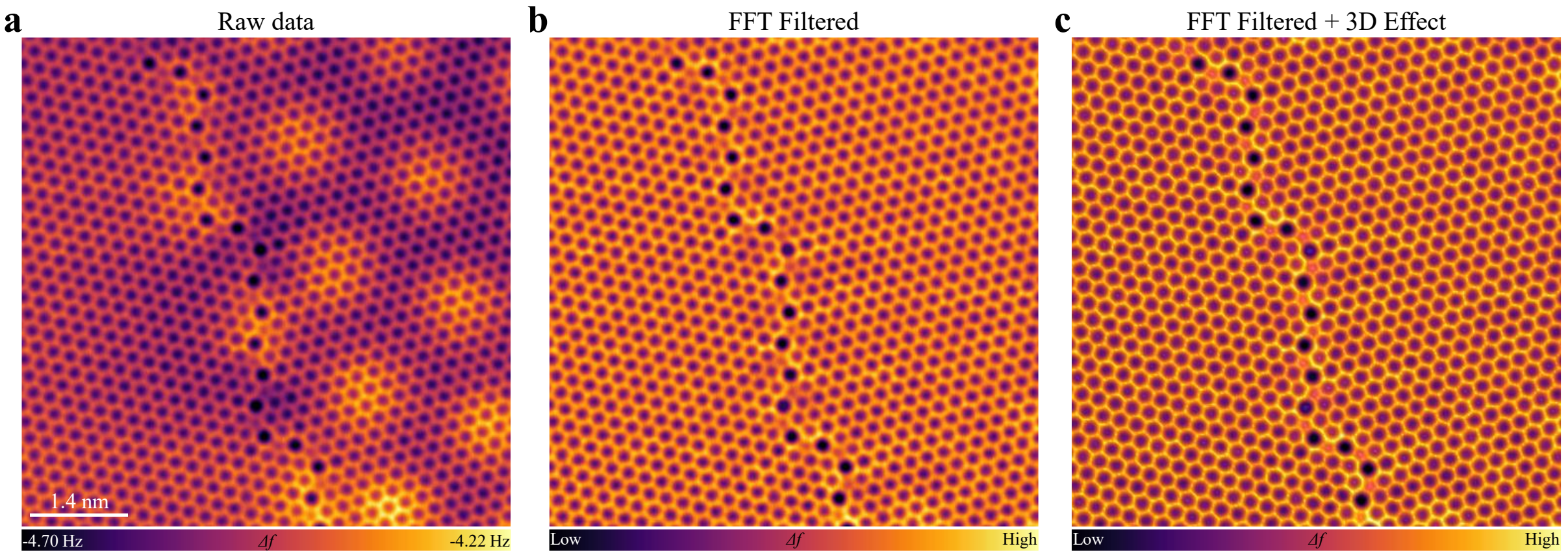

**Figure S2. Example of data processing applied to the constant height ncAFM images**. The experimentally measured raw data is shown in (a). By filtering out the low frequency signals of this image, we obtain the image shown in (b). By further applying a 3D effect, that is, we illuminate the image in the z-axis to enhance the visibility of the chemical bonds, we obtain the image depicted in (c).

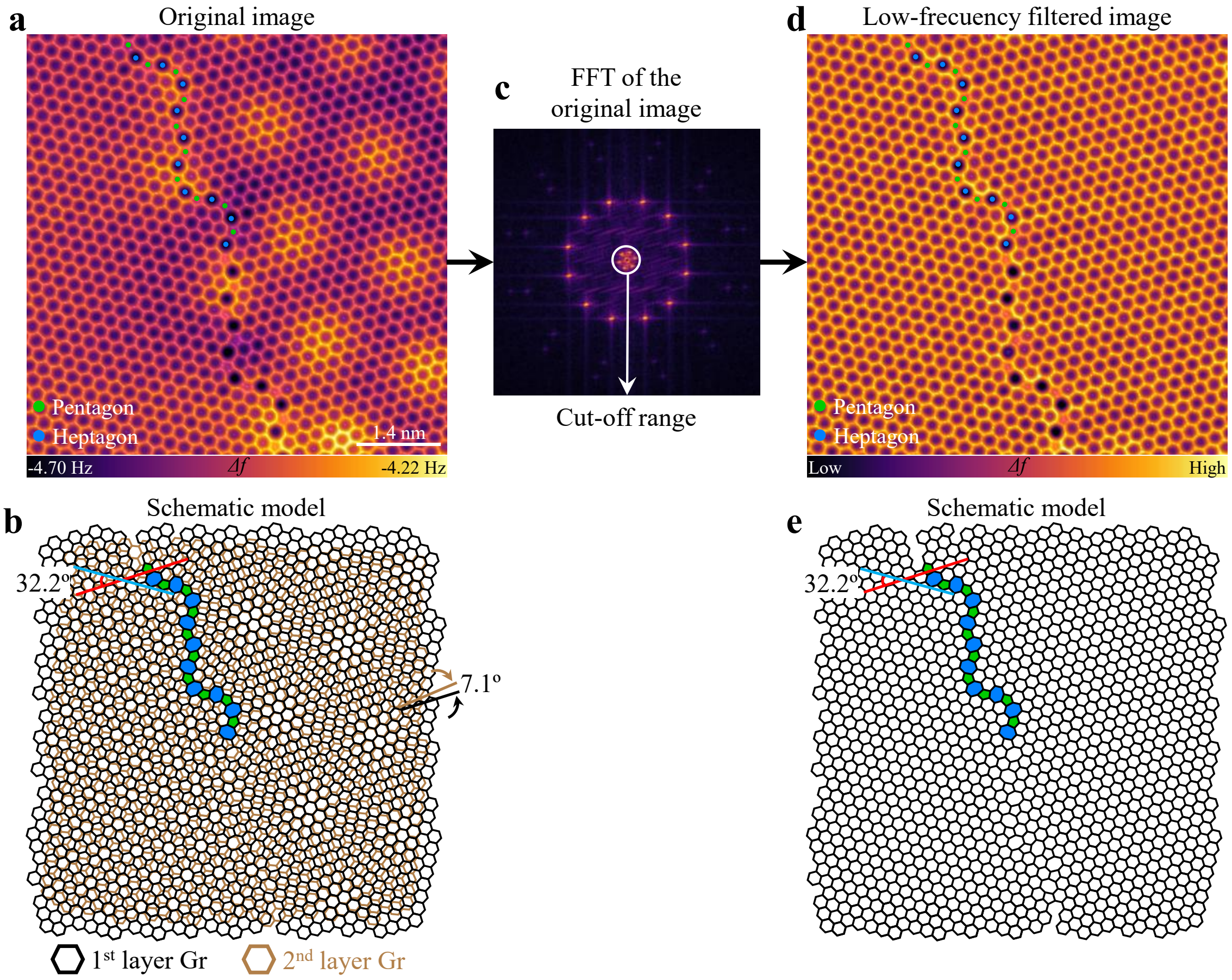

**Figure S3. FFT filtering procedure applied to the ncAFM images.** a) Original measured constant height ncAFM image. b) Schematic representation obtained by placing a graphene sheet, that features two rotational domains with a misorientation angle of 32.2º, on top of another graphene sheet. Both graphene layers are rotated by an angle of 7.1º with respect to the upper-right domain. This schematic model reproduces all the main characteristics depicted in (a). c) Calculated FFT pattern of the image shown in (a). Using this pattern as reference, we filtered the original image by cutting off all the signals within the white circles (low frequency signals). The result is depicted in (d), which corresponds to the image shown in Figure 1a in the main text. e) Schematic representation that reproduces the characteristics of the image (d). Therefore, our FFT filtering procedure eliminates the signal contribution coming from the bottom graphene layer, so that we can focus our attention only on the topmost graphene layer. A 3D effect has been applied to all images.

## Supplementary note 2: Theoretically modelling and simulated ncAFM images of graphene grain boundary

To account for the experimental observations presented in the main text we performed theoretical modeling of graphene grain boundary. We employed two different grain boundaries models, as illustrated in Figure S4, namely the ideal relaxed GB and the strained and corrugated GB. In the case of the ideal GB (Figure S4a), we constructed it by merging two graphene sheets that present a misorientation angle of exactly 32.2º between them. The employed supercell has a size of 2.797×0.89 $nm^2$ and with a total number of C atoms of 96. As can be seen from the side view, the ideal GB remains in a fully flat configuration. Conversely, in the case of the strained and corrugated GB, we used the same base model as for the ideal GB, but additionally imposed a compressive uniaxial strain ($s$) along the x-axis, concentrated within the GB region, and explicitly enforced a vertical corrugation ($h$) (see Methods)

Using the above-described models, we simulated the corresponding ncAFM images using a CO molecular terminated tip-apex (see Methods section in the main text) and considering the potential impact or not of the mobility of the CO molecule on the tip-apex. For this purpose, we mimic a static CO molecular tip or consider its relaxation effect ($k$ = 1.5 eV/$rad^2$), where the torsional spring constant $k$ is intentionally set much higher than typical values used for modeling real CO-functionalized tips, in order to limit the probe lateral relaxation as expected for the oxidized apexes used in these experiments. They obtained simulated ncAFM images of the ideal and strained and corrugated GBs are shown in Figure S5 at different tip-sample distances ($Z_{ts}$). The theoretical images used for comparison with the experiments correspond to the simulated ncAFM images at $Z_{ts}$ = 297 pm and considering the relaxation effect of the CO tip. The choice of this intermediate tip-distance as our reference relays in the fact that all the experimental constant height ncAFM images were recorded always in the attractive short-range force regime. Additionally, we preferred to opt for considering the CO tip relaxation effect in the simulated images as it gives qualitatively the best fit results with our experimental images, although we emphasize that, quantitatively, the difference with respect to a static tip is almost undetectable.

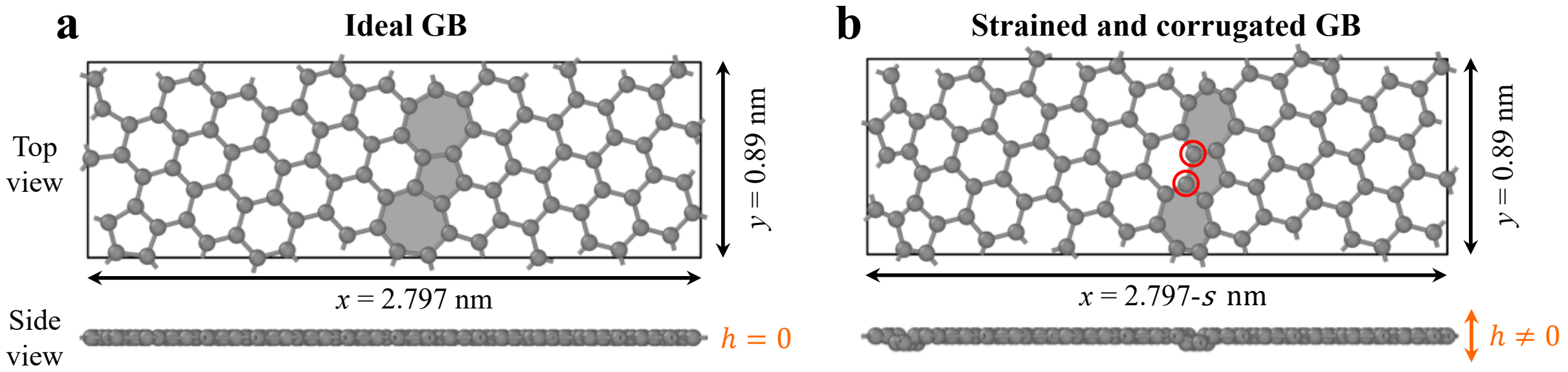

**Figure S4. Grain boundary supercell used for the simulation of the ncAFM images.** a) Supercell used to simulate the ideal, fully flat GB configuration. b) Supercell used to simulate a strained and corrugated GB, which is built from the ideal GB by rigidly displacing one grain toward the other by a distance $s$, while simultaneously reducing the supercell dimension along the $x$-direction by the same amount and by enforcing a vertical corrugation ($h$) in the two C atoms shared by the 5-6 pair flanked by two 7 rings (highlighted by the red cycles).

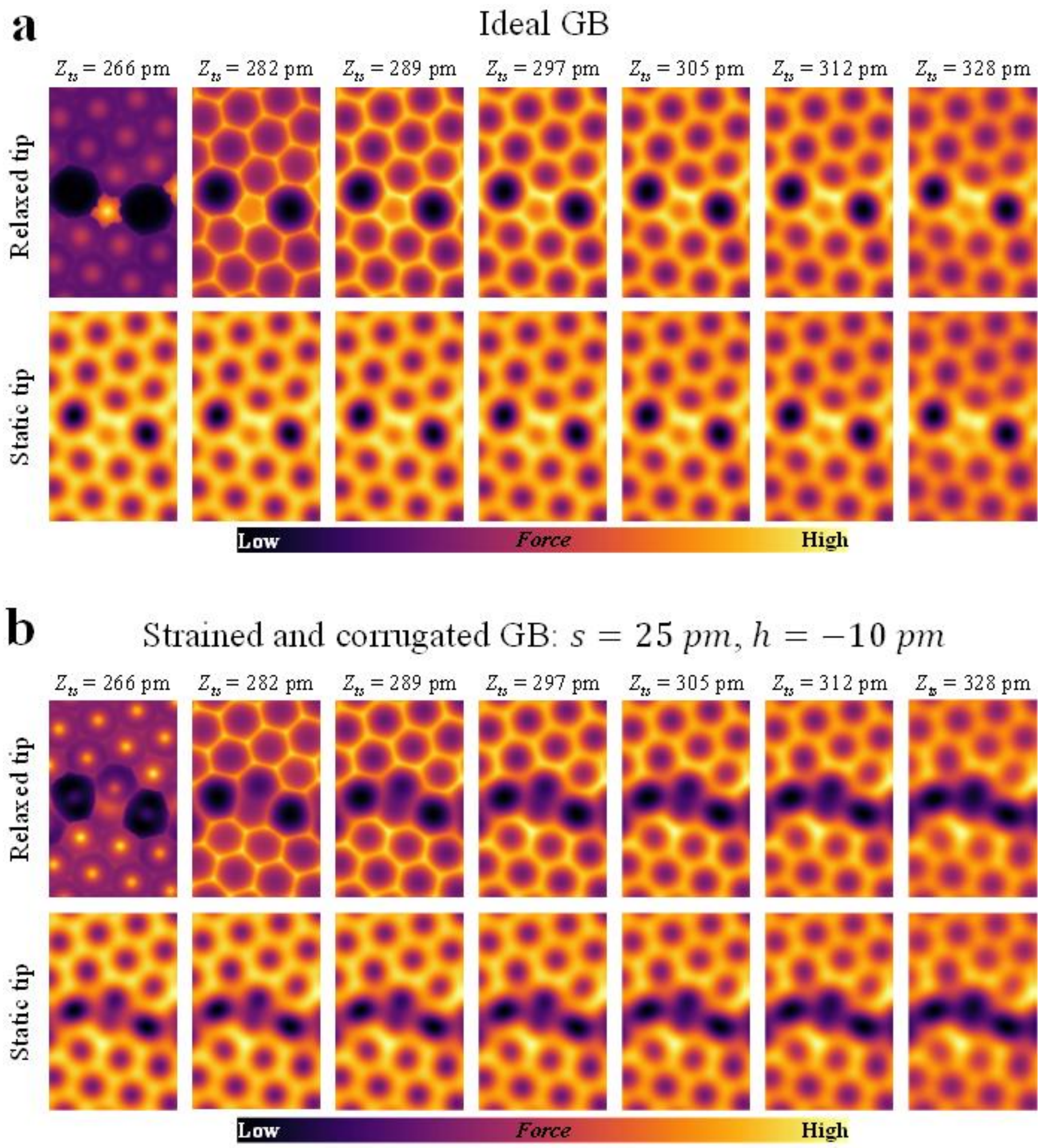

**Figure S5.** Simulated ncAFM images of an ideal GB (a) and a strained and corrugated GB (b) at different tip-sample distance ($Z_{ts}$). A CO tip was employed to simulate all the images. Also, we show, comparatively, images simulated considering the effect of tip relaxation ($k$ = 1.5 eV/rad$^2$) or not (static).

In the experimental ncAFM images of strained grain boundaries, two atoms at the edge of one grain appear to be missing. This apparent absence can be explained by considering the simulated structural relaxation: the two atoms relax downward by approximately 10 pm. This vertical displacement has two concurrent effects on the ncAFM contrast (see Figure S6). First, the reduction in height directly lowers the local tip-sample Pauli repulsion, decreasing the *frequency shift* signal. Second, an increased out-of-plane localization of the corresponding C–C bond leads to a more confined electronic density, further reducing the spatial extent of the short-range interaction. As a result, these atoms contribute less contrast and may appear absent in the images.

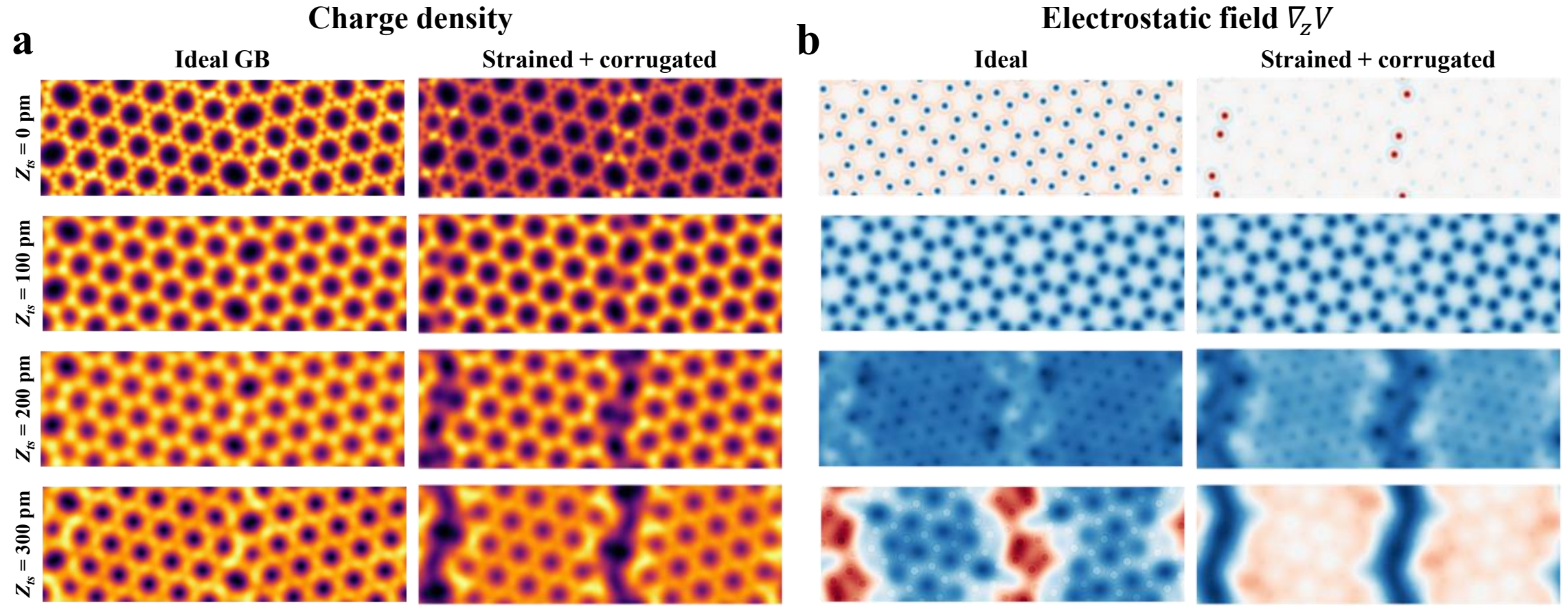

**Figure S6. Simulated charge density and electrostatic field maps for ideal and strained grain boundaries.** (a) Charge density maps computed at different heights $Z = 0$, 100, 200, and 300 pm for the ideal GB (left column) and the strained and corrugated GB model (right column). (b) Corresponding vertical component of the electrostatic field, $\nabla_Z V$, maps projected at the same heights for the two GB models. While the ideal GB maintains symmetric and uniform electronic features across the interface, the strained and corrugated GB exhibits a pronounced modulation of the charge distribution and the $\nabla_Z V$, particularly along the GB line, with signs of symmetry breaking and local bond rearrangements. The two boundary atoms that have been displaced downward exhibit negligible charge density at the highest tip–sample distances, due both to their vertical displacement and to the increased out-of-plane localization of their bond.

**Figure S7. Simulated ncAFM contrast as a function of the corrugation.** Simulated ncAFM images of a non-corrugated GB (left column) and corrugated GB for -20 pm, -10 pm and +10 pm out-of-plane displacements (minus/plus means downward/upward displacement) at different tip-sample distance ($Z_{ts}$). All the images correspond to an imposed compressive in-plane strain of 25 pm. The contrast in the boundary is very dependent on the corrugation. Best agreement with the experiment is found for a downward corrugation of ~ 10 pm, where features originated by the bonds of the 5-7 rings are still barely observed while the contrast on the bond between the two downward displaced atoms is missed.

**Supplementary note 3: Manipulation of the metastable graphene grain boundaries**

In the main text, we discussed that the observed atomic structure of the metastable graphene grain boundary can undergo modification through local manipulation (e.g., see Figure S8). In this work, such manipulation procedure consisted of repeated indentations using the ncAFM tip at different places within the GB area and around its close vicinity with the feedback loop open and maintaining the cantilever oscillation active during the whole indentation process. This indentation consisted of performing frequency shift *versus* distance (*Δf(z)*) curves, with typically recorded ones plotted in Figure S9 as reference. As seen, the tip-sample interaction extends well beyond the repulsive regime—likely making atomic contact—where the scanner piezo moves an average of 4-6 nm further compared to normal imaging tip-sample distances (~ 1-5 Å). This suggests that the energy exchange between the tip and sample is probably sufficiently high to facilitate surface relaxation of the GB region. As a result, the GB transitions from a strained and corrugated state to a fully flattened ideal configuration. In general, the modification does not occur gradually, but we believe in a sudden way during one of the indentations, as after repeated indentations a change in the GB area is observed, where regular pentagon and heptagon rings emerge. Once it is modified, it is impossible to reverse back to a metastable configuration, as far as we have tried during the experiments by repeating the very same manipulation procedure, although higher statistics are needed to confirm this. The existence of potential modification of the graphene grains is not observed (e.g. twist angle modification).

An example between the initial and final stage of the grain boundary region, at different fields of views, after the manipulation is shown in Figure S10, where one can clearly observe differences in the atomic structure in the GB area. Additionally, we note that these modifications are also appreciable even in the experimentally acquired raw ncAFM images, as evidenced in Figure S11.

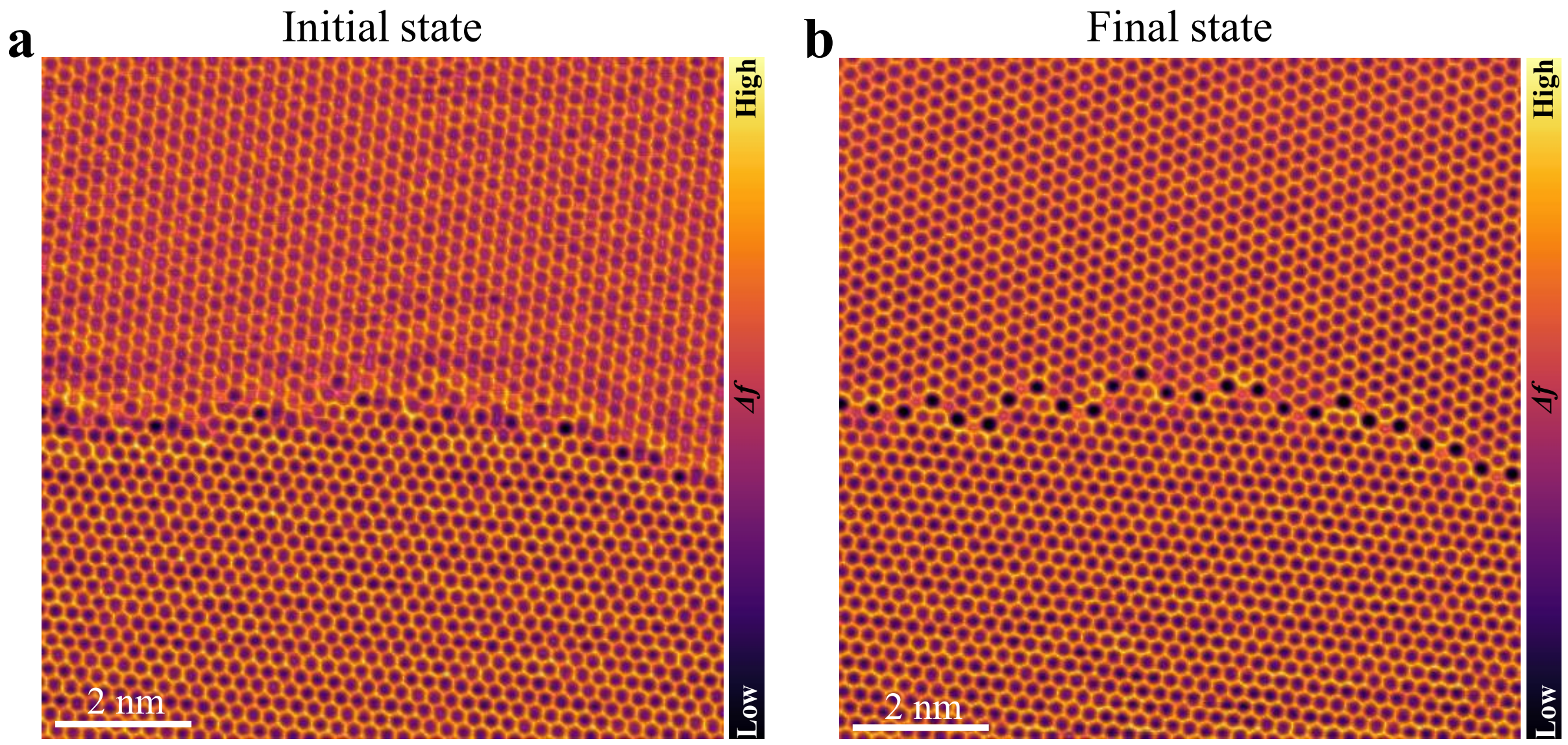

**Figure S8.** a,b) Corresponding enlarged view of ncAFM images shown in Figure 2a,e, revealing the GB region before and after ncAFM tip manipulation. FFT filtering and 3D effect have been applied to all the images.

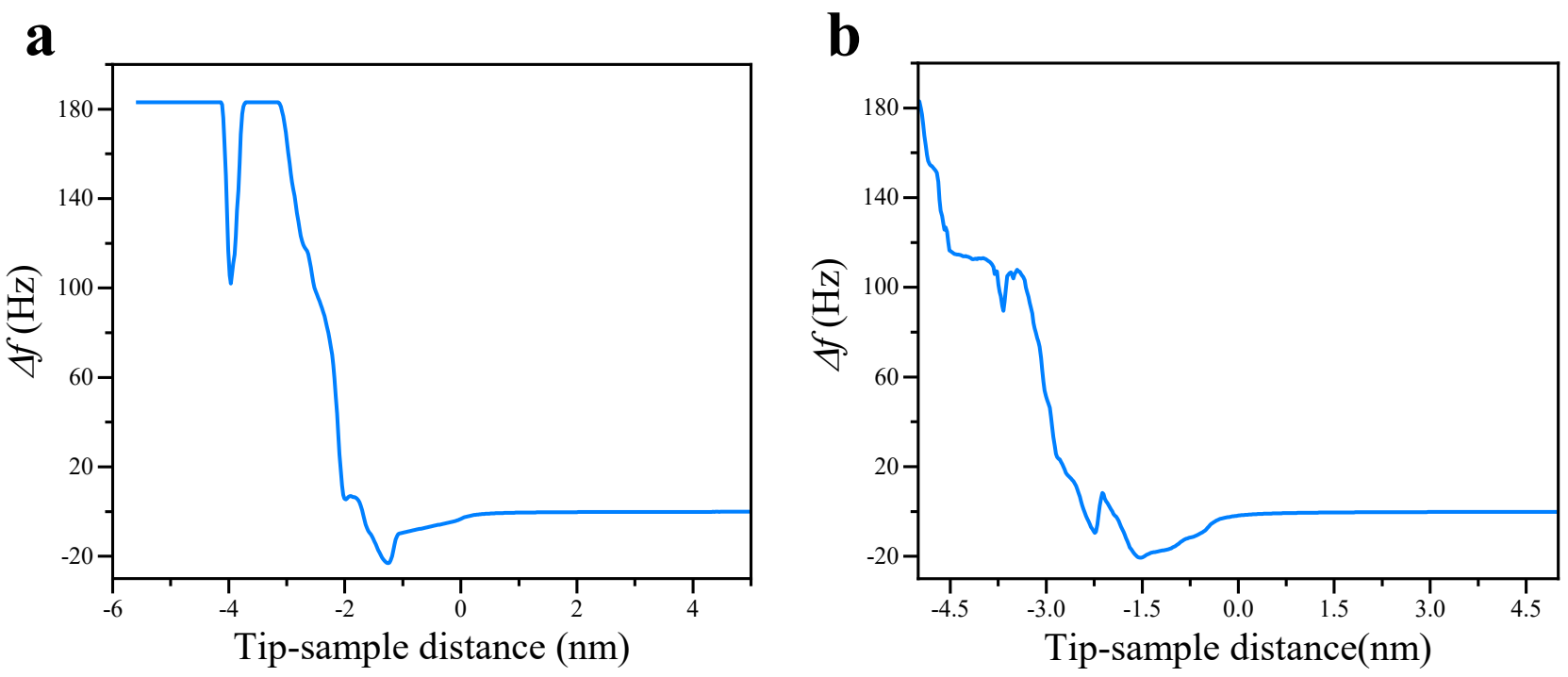

**Figure S9.** Typical $\Delta f$ curves registered as function of the tip-sample distance during the application of indentation on the GB.

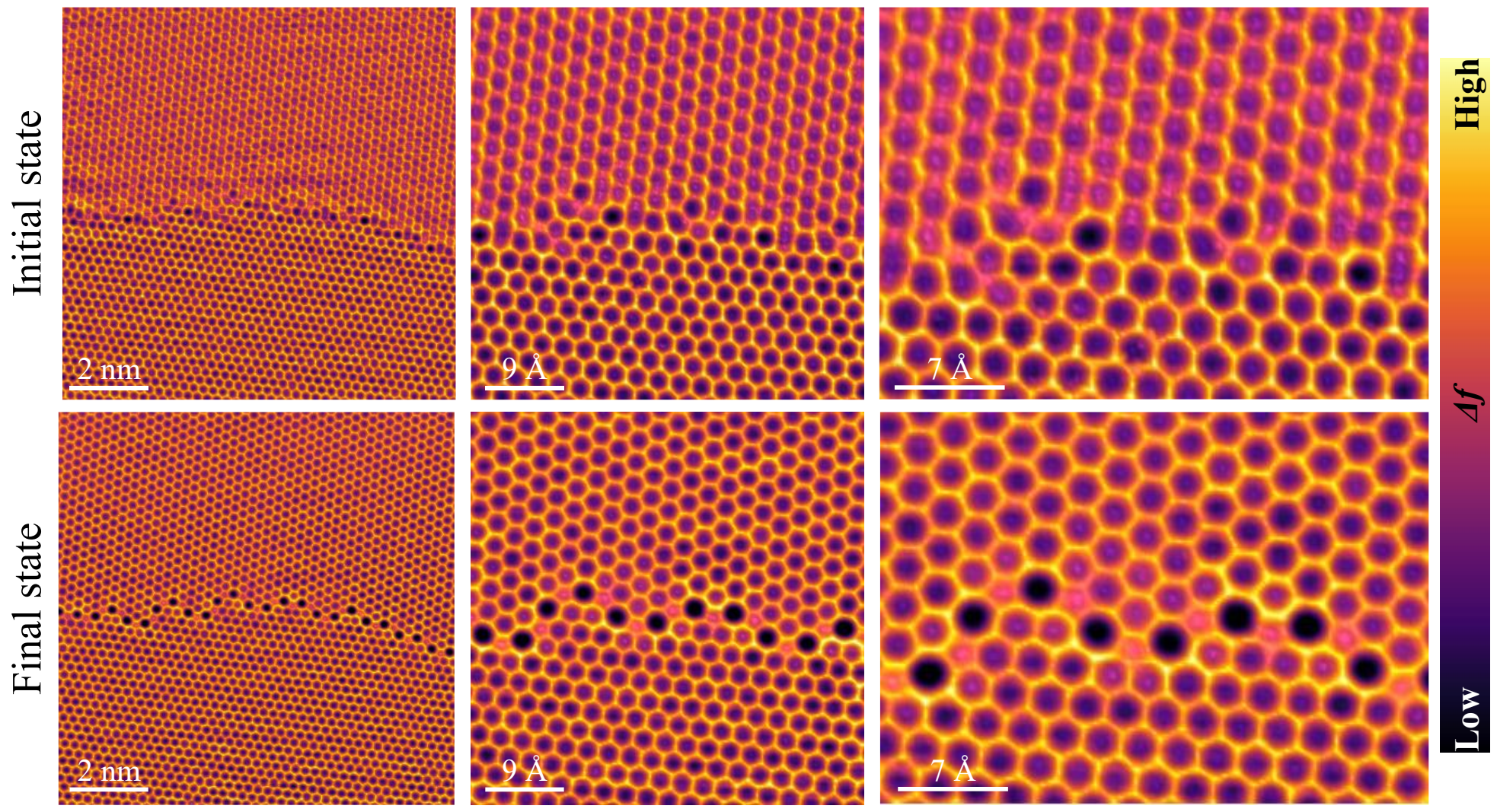

**Figure S10.** Different magnification constant height ncAFM images of the GB region described in Figure 2 before and after manipulation with the ncAFM tip. FFT filtering and 3D effect have been applied to all the images. FFT filtering and 3D effect have been applied to all the images.

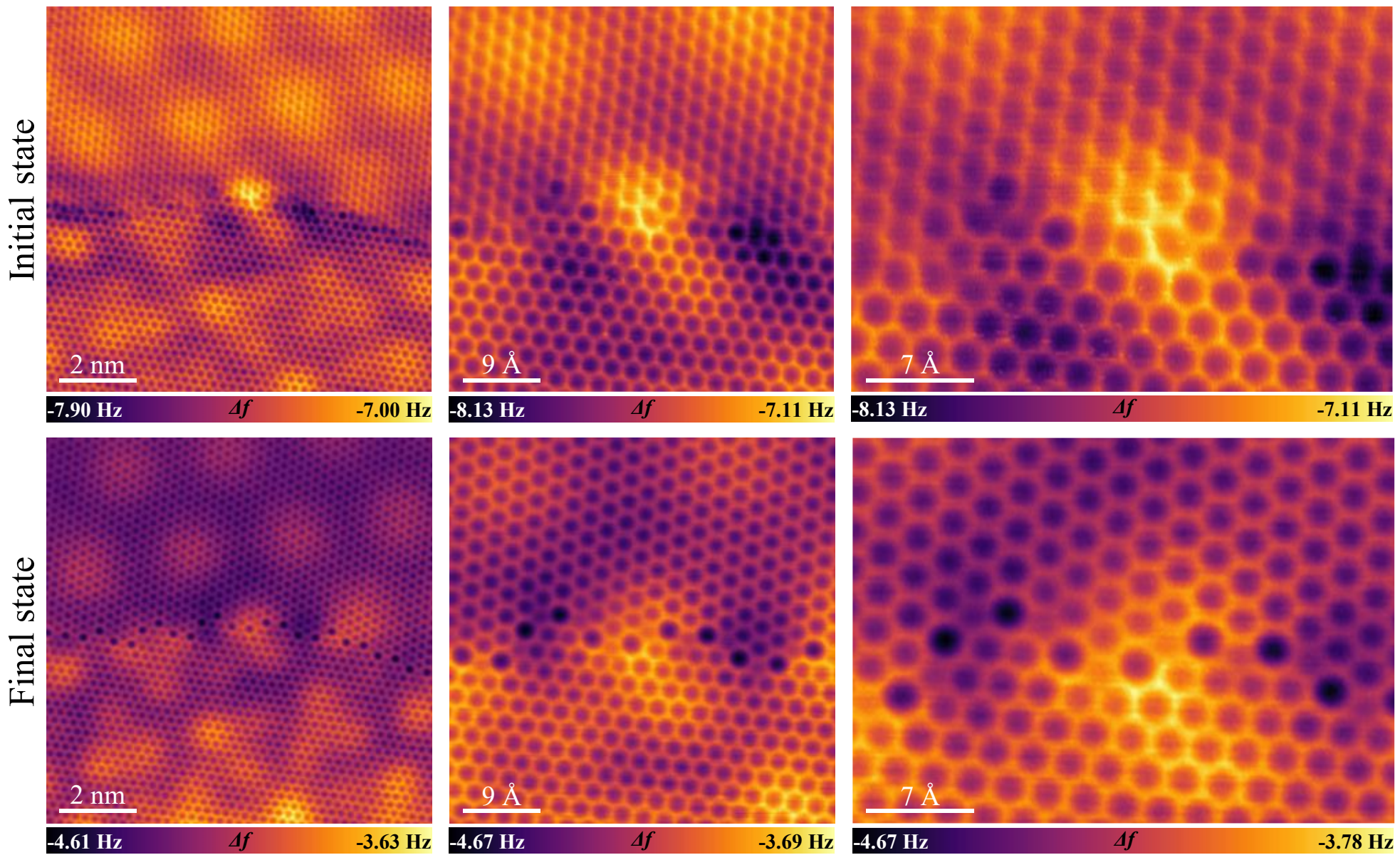

**Figure S11.** Experimentally measured constant height ncAFM images at different fields of view of the GB described in Figure 2 before and after manipulation with the ncAFM tip. These images are the experimental raw (without applying the filtering process) images of the one shown in Figure S10.

**Supplementary note 4: Additional data regarding the coexistence of metastable and ideal GBs, and their manipulation**

To further support the existence of both energetically metastable and ideal grain boundaries on the as-grown samples of graphene, in Figure S12 we show additional examples of these two types of GBs that present distinctive geometrical features along the GB area. In the same line, supporting example of the manipulation of metastable GB driven into a more stable and ideal configuration is provided in Figure S13. However, in this case, we did not determine the exact local corrugation and strain needed in the formation of the initial metastable GB.

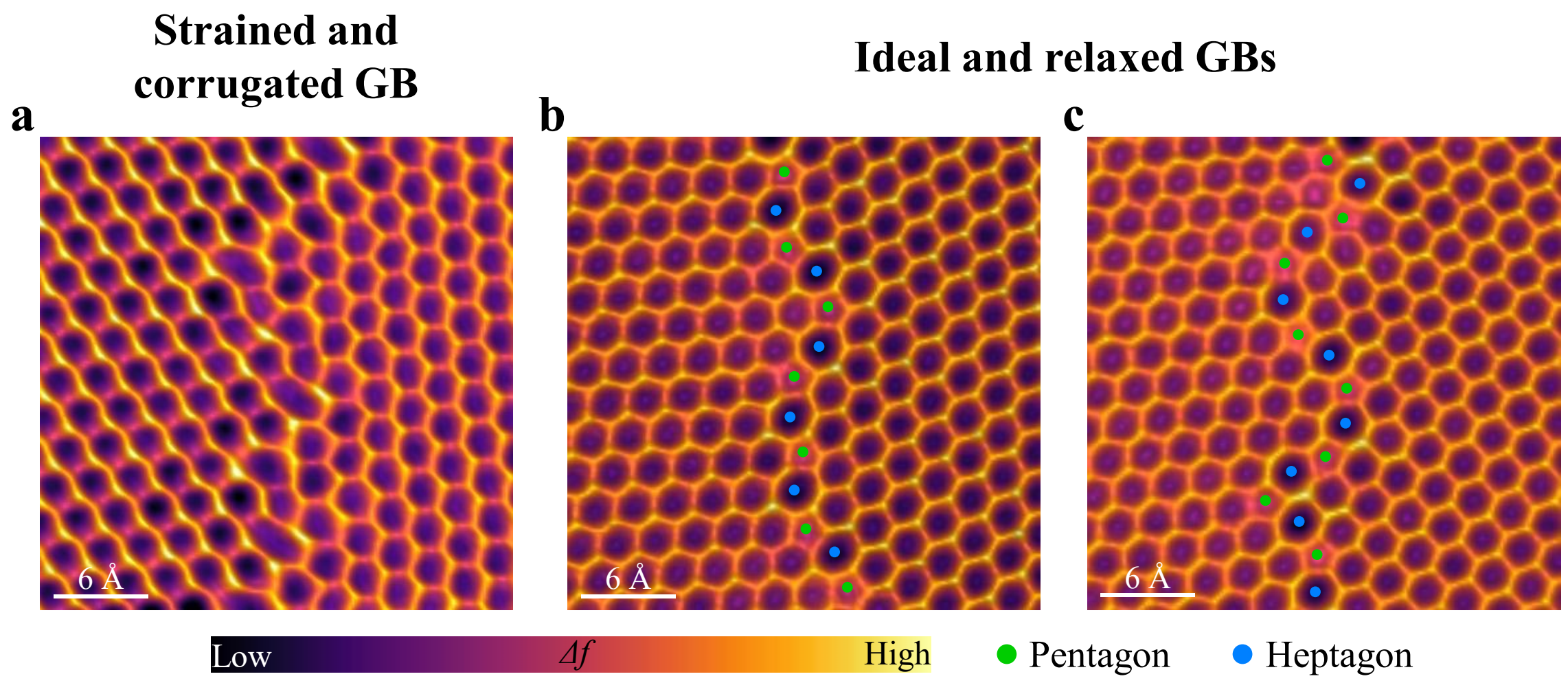

**Figure S12. Additional data of strained and corrugated and ideal GBs.** Representative constant height ncAFM images recorded at different sample regions showing complementary examples of strained and corrugated GB (a), as well as ideal GBs (b-c). FFT filtering and 3D effect have been applied to all the images.

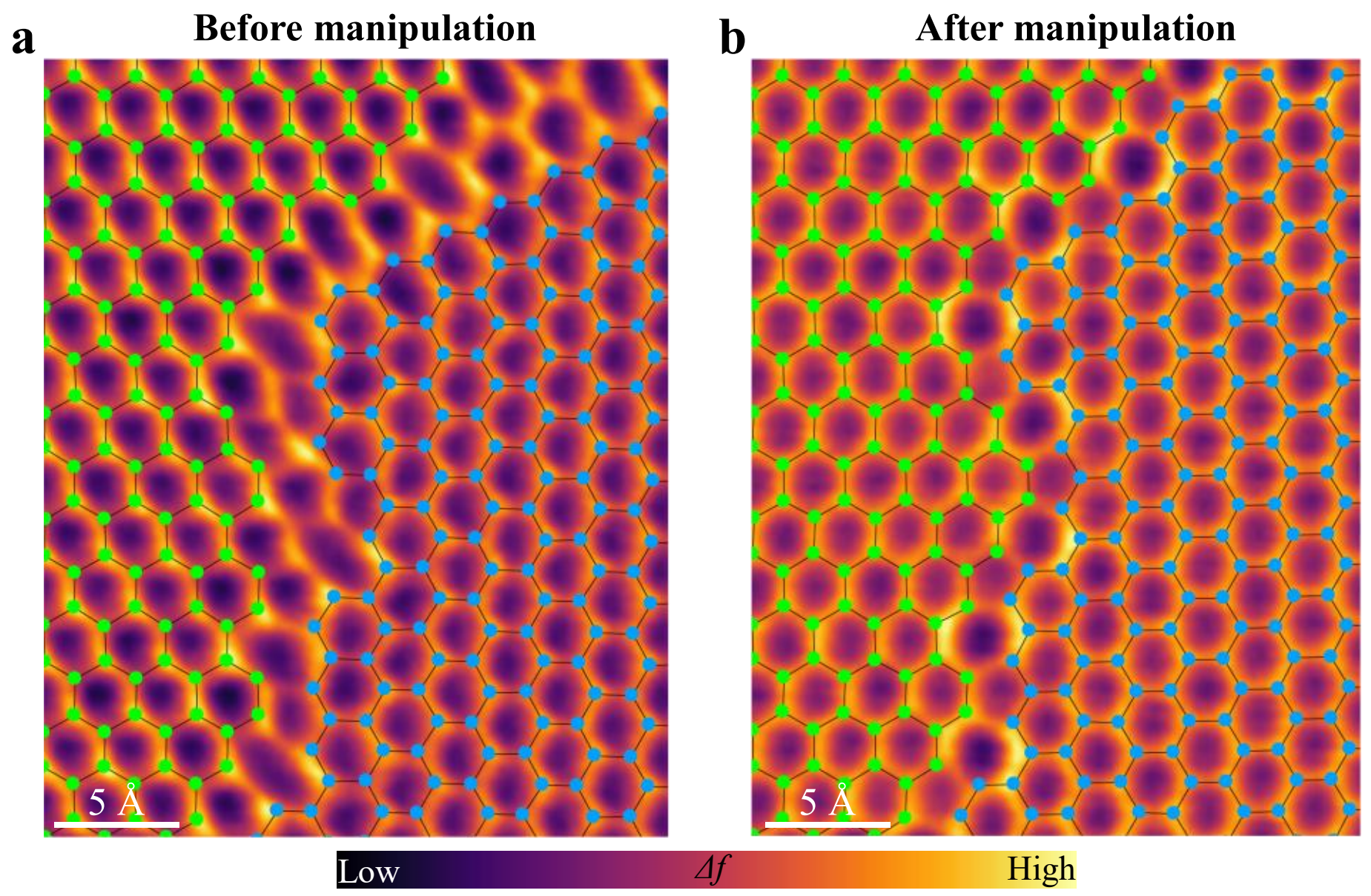

**Figure S13.** Further example of tip-induced manipulation of an initial metastable GB (a) driven to a stable and regular GB (b). FFT filtering has been applied to the images.

**Supplementary note 5: Experimental evidence of strain in the metastable grain boundary**

To experimentally sustain the presence of strain in the GB area, we superimposed the acquired ncAFM images—captured before and after ncAFM tip manipulation—onto theoretical models of the GBs, comparing strained and corrugated structures against flat and planar configurations, as illustrated in Figure S14. In both scenarios, our experimental observations exhibit a remarkable alignment with the corresponding theoretical models, reinforcing the validity of our conclusions. Additionally, we experimentally observe that the structural distortions in the metastable GBs area impact also the graphene structure over several nanometers' length scale. This is evidenced by the observed "quinoidal" configuration of the graphene honeycomb in the upper grain (see Figure S14a), where 2 out of 6 bonds are imaged which brighter contrast. This is indicative of potential differences in the C bond lengths.

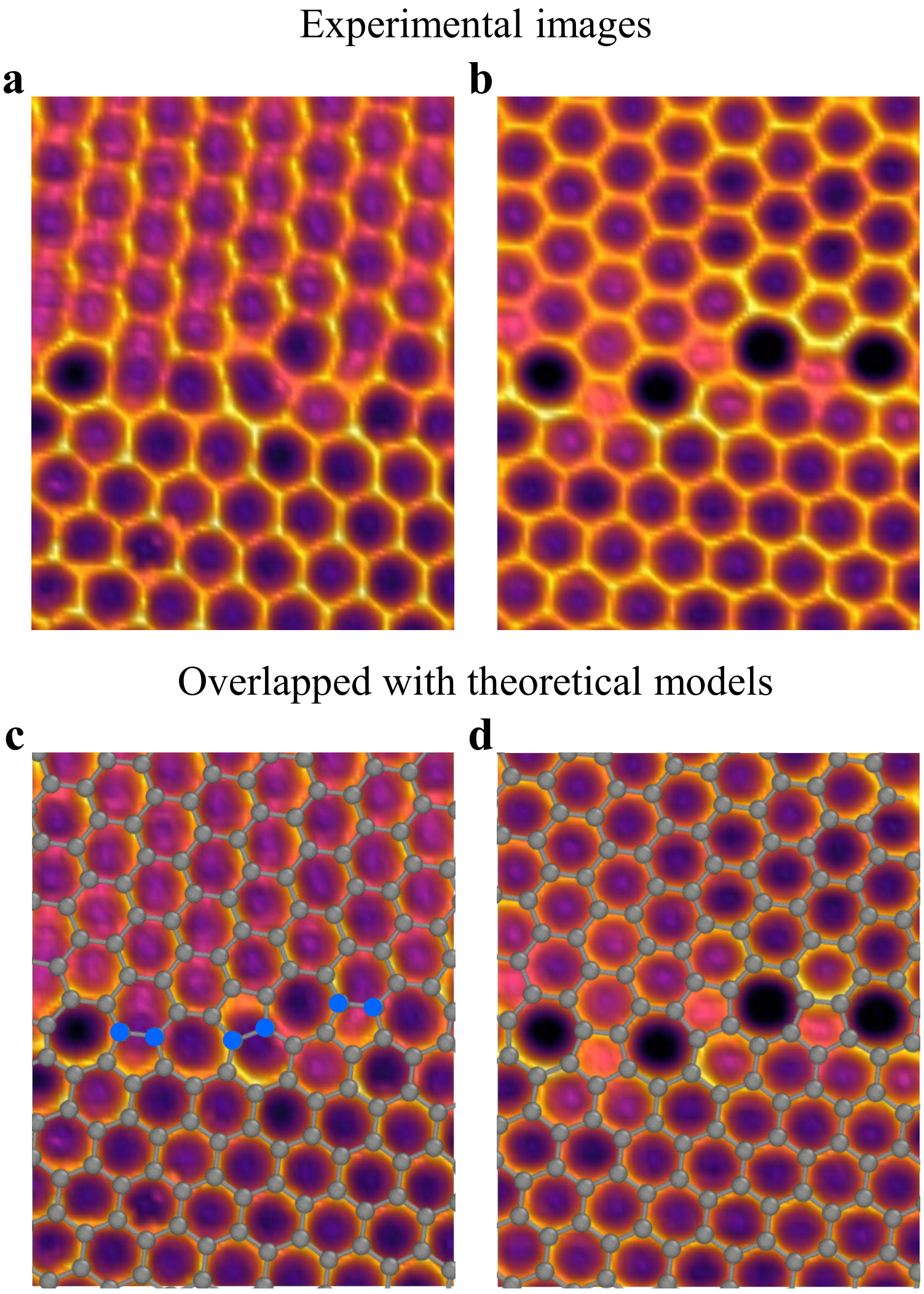

**Figure S14.** Comparison between the experimental acquired constant height ncAFM images (a,b) and their respective best theoretical GB model superimposed on them (c,d). FFT filtering and 3D effect have been applied to all the images. In (c), the overlapped model corresponds to a GB under the effects of compressive strain $s$ = 25 pm and vertical corrugation $h$ = -10 pm (lowered). Blue dots highlight the C atoms in the boundary that are under the compressive strain effect and vertical corrugation. Meanwhile, in (d), the model corresponds to an ideal GB (flat and planar configuration). These are the models used to simulate the ncAFM images depicted in Figure 2g,h.

**Supplementary note 6: Morphology of the metastable grain boundary area**

The induction of strain in the GB can likely be attributed to the presence of structural defects on the surface near the GB area. As shown in Figure S15, several prominent defects, such as wrinkles and beads are observed in proximity to the GB (highlighted by green squares). These defects may serve as pinning points, preventing the complete relaxation of the metastable GB structure and consequently giving rise to the asymmetric geometric arrangement of pentagonal and heptagonal rings, as discussed in Figure 2 of the main text. However, upon manipulation with the ncAFM tip, this strain is effectively erased, allowing the GB to revert to its fully relaxed planar configuration.

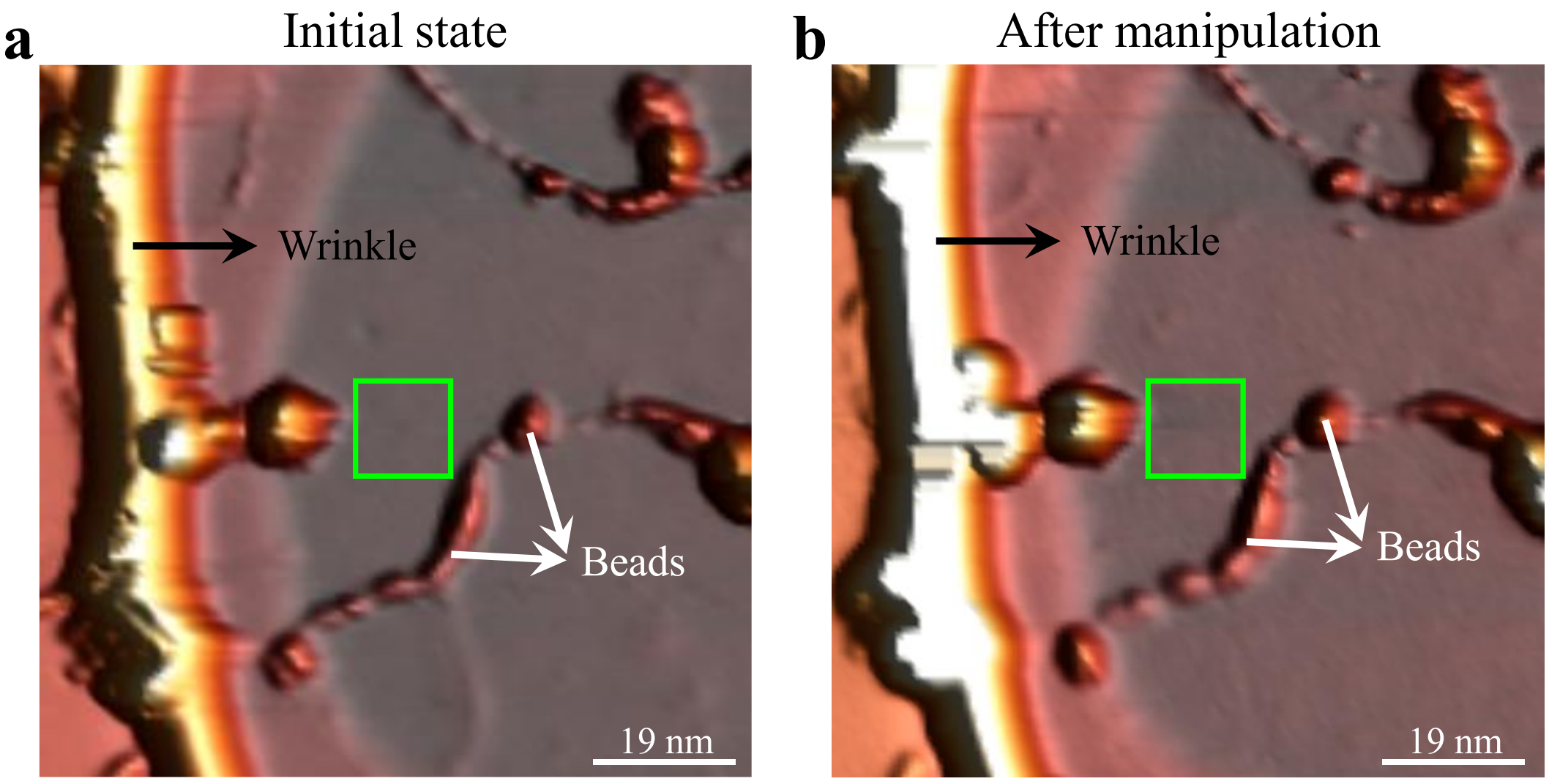

**Figure S15.** Large-scale topographic ncAFM images showing the morphology of the region around the GB (indicated by the green squares) before and after we performed the manipulation described in the main text. Surface defects such as wrinkles and beads are highlighted. A derivative filter has been applied to all images to enhance the features. Acquisition parameters: a) $\Delta f$ = -1.5 Hz, $V_{CPD}$ = 0 V. b) Δf = -1.5 Hz, $V_{CPD}$ = 0 V.

**Supplementary note 7: Long range perturbation and Clar's sextet theory in graphene grain boundaries**

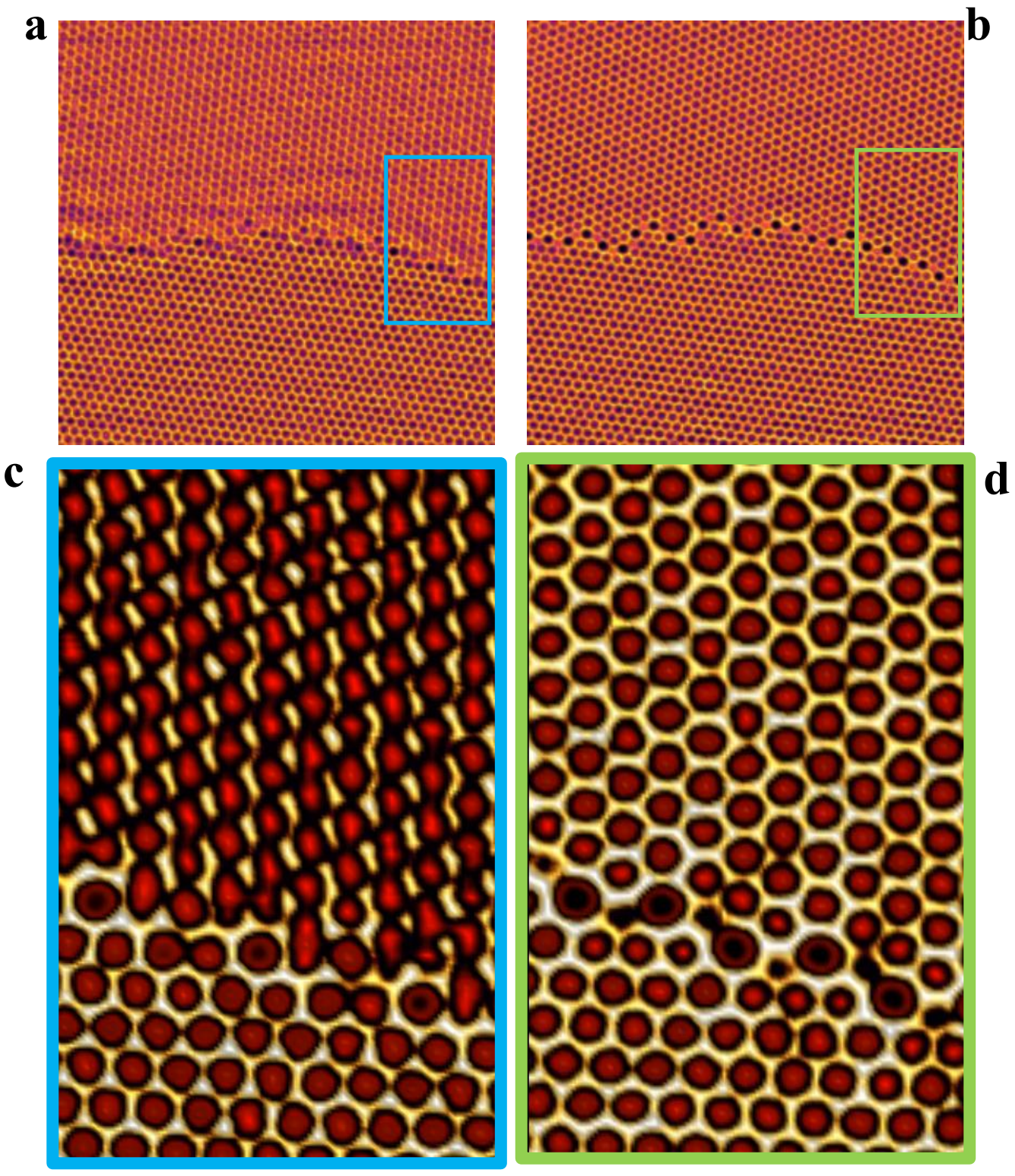

**Figure S16.** a,b) Same ncAFM images as Figures 2a,e of the main manuscript, acquired in the same region before and after manipulating a strained metastable GB. c,d) Zoom in of the region outlined by a blue (green) rectangle. Image contrast has been adjusted to better visualize the bond structure. In the strained GB (a,c), the upper grain displays a peculiar quinoidal bond texture in which one bond per carbon hexagon has higher intensity. On the contrary, in the relaxed grain boundary (b,d), this so-called quinoidal bond texture is not seen, and both graphene domains adopt an ideal graphene appearance

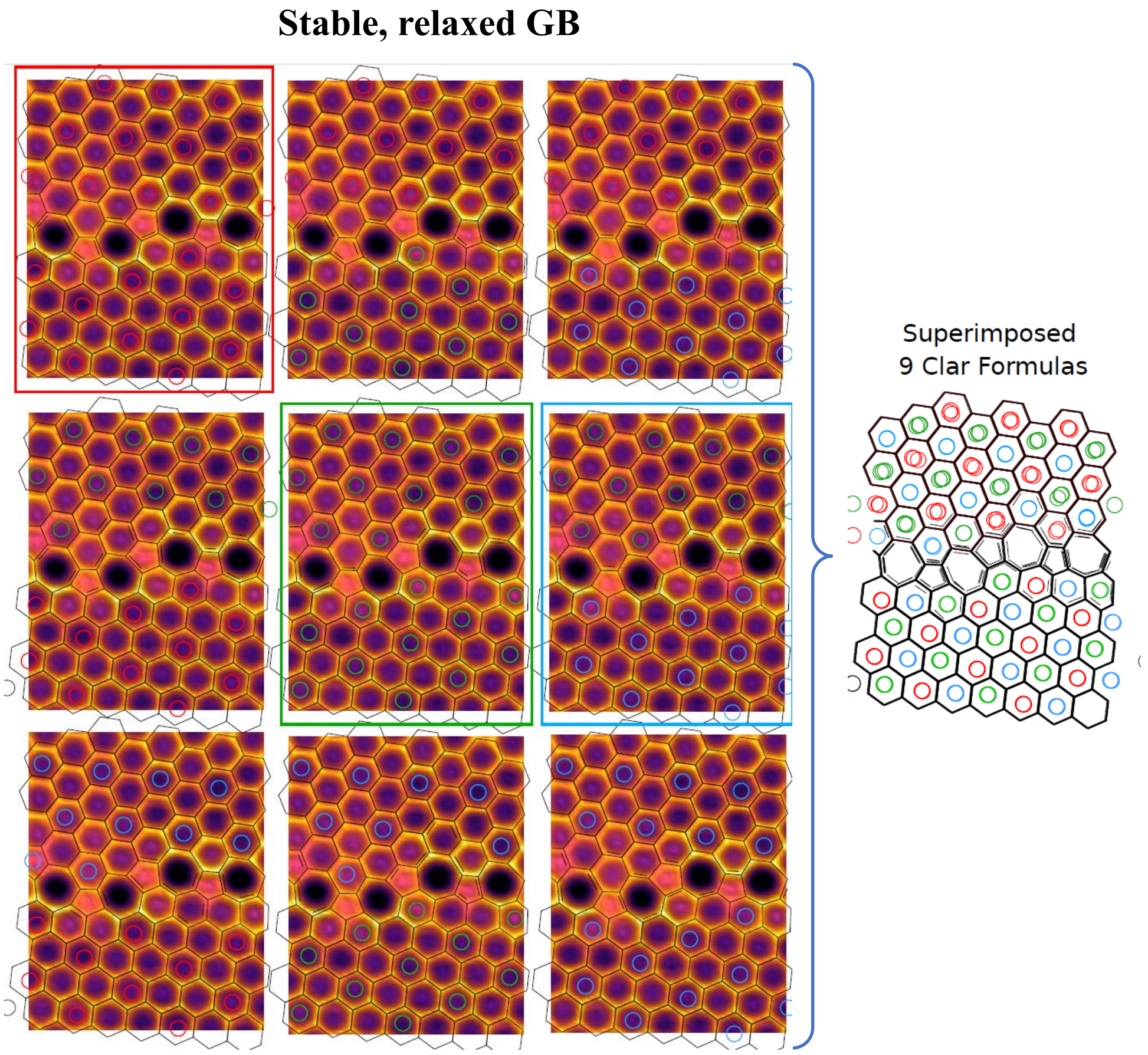

**Figure S17.** For the relaxed, stable grain boundary, both graphene domains adopt an all-benzenoid configuration—an ideal graphene appearance. Clar formulas that preserve the threefold degeneracy of Clar structures on either side of the boundary can be constructed, provided that the double bonds formed during grain-boundary creation remain confined within it.

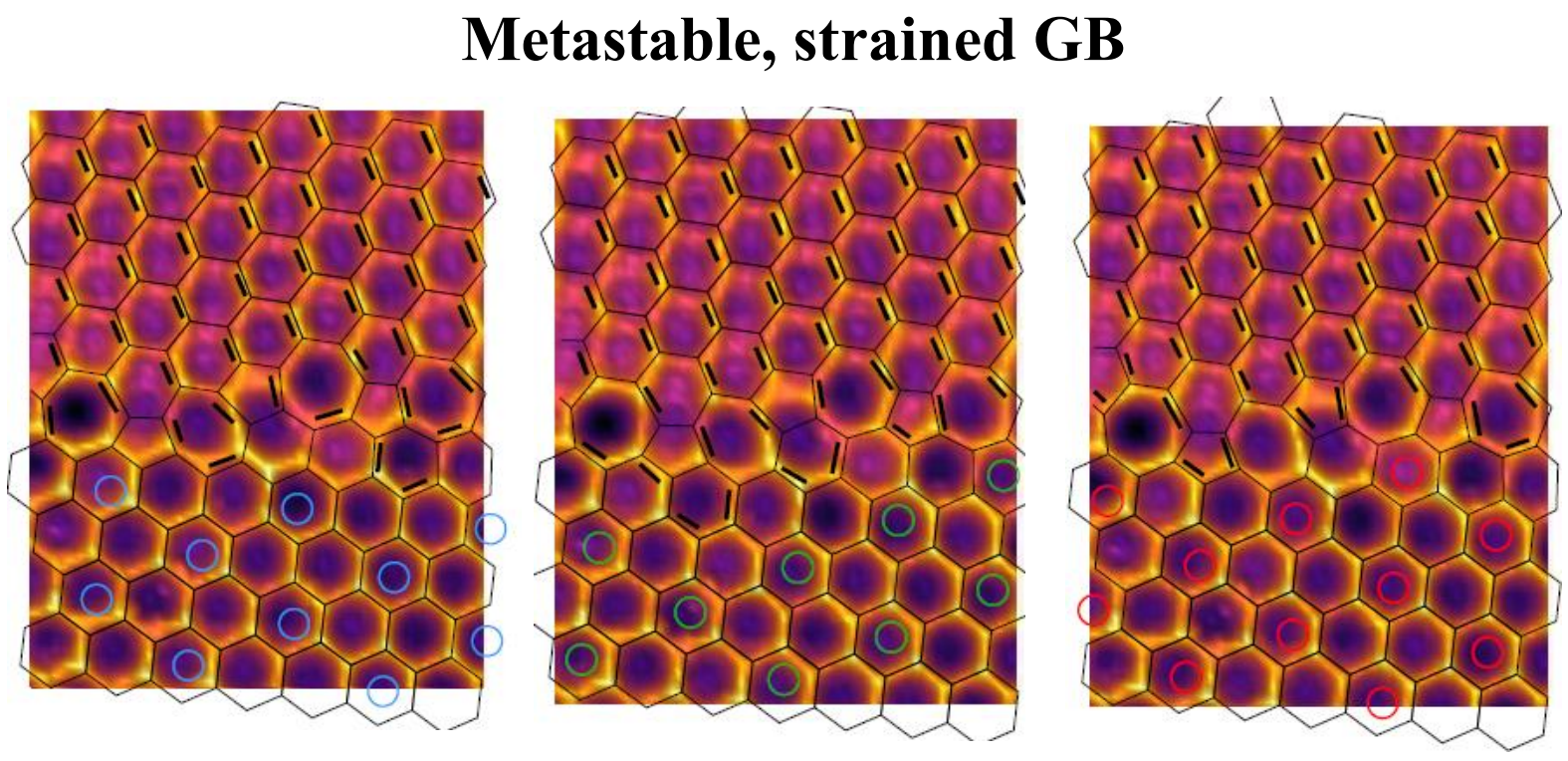

**Figure S18.** Clar formulas can be constructed to reproduce the quinoidal bond pattern observed in the upper grain. However, our current understanding of the grain boundary details remains insufficient to explain why this specific configuration is favored over other possible arrangements.

**Supplementary note 8: Moiré patterns on Gr/SiC(000-1)**

As mentioned in the Methods section, our experiments were conducted on polycrystalline graphene surfaces grown on SiC(000-1). This growth method results in the formation of multilayer graphene with significant vertical rotational disorder, which facilitates electronic decoupling between consecutive layers. Additionally, this rotational misalignment leads to the emergence of multiple moiré patterns on the surface, characterized by considerable heterogeneity in their periodic wavelengths. In Figure S19 we show several representative moiré patterns observed in the present work on the graphene/SiC(000-1) surfaces, measured using either constant *frequency shift* ($\Delta f$) or constant height ncAFM acquisition modes.

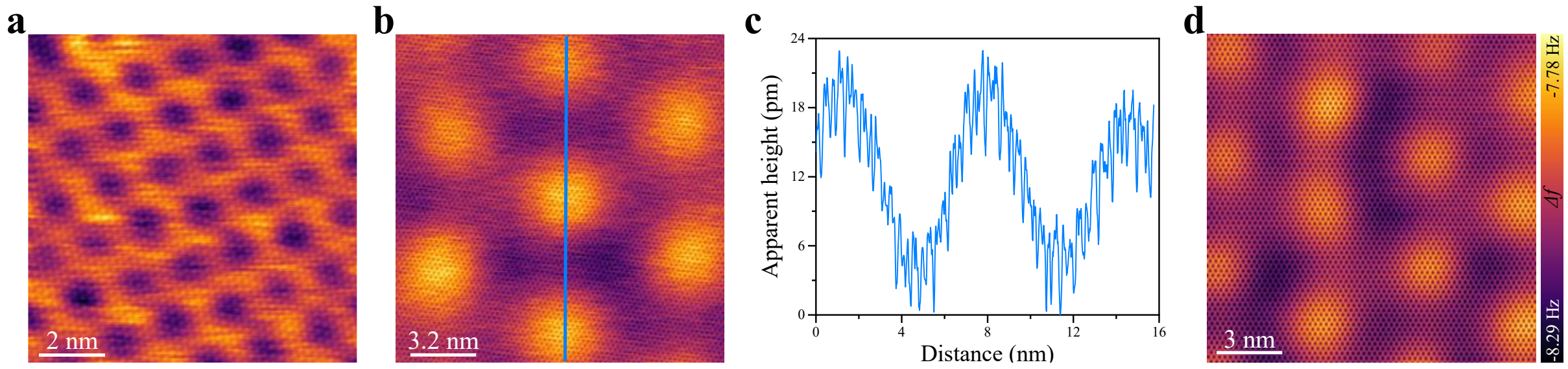

**Figure S19. Graphene moiré patterns observed on the surface of Gr/SiC(000-1) by ncAFM.** a-b) Atomically resolved constant *frequency shift* ncAFM images of different moiré patterns. c) Apparent height profile along the blue line drawn in (b). Despite its large periodicity (~ 7 nm), the moiré pattern shows a small apparent corrugation (~ 12 pm). d) Atomically resolved constant height ncAFM image of large periodicity moiré pattern. Acquisition parameters: a) $\Delta f$ = -4 Hz, $V_{CPD}$ = -40 mV. b) $\Delta f$ = -11 Hz, $V_{CPD}$ = 50 mV. d) $V_{CPD}$ = 0 V.

**Supplementary note 9: Conditioning of the experimental ncAFM tip-apex**

In this note we detail how we carried out the conditioning of our tip-apex before performing high-resolution imaging of the atomic structure of graphene grain boundaries with our ncAFM. In Figure S20 we show two representative ncAFM topographic images revealing the overall surface morphology of the Gr/SiC(000-1) samples studied in this work. We observe the existence of several surface imperfections, such as structural defects like wrinkles and beads, and aggregates of contaminants, whose chemical composition is unknown. Indeed, we prepare our tip-apex on top of these contaminants by performing indentation (*Δf(z)* spectroscopy) or continuous scanning on top of these aggregates until a sudden enhancement of the spatial resolution is observed in topography images. Then, we move the tip to the preselected GB areas (like those highlighted in Figure S20) to carry out the high-resolution imaging of their atomic structure. Although we don't know the exact chemical nature of our tip-apex, we note that this procedure has allowed us to measure all the bond-resolved ncAFM images shown in this work routinely. Therefore, this procedure is empirically reliable and our simulations using a proxy CO tip reproduce the experimental contrast. However, we highlight that there could be some experimental constraints in the future when these unknown contaminants are not present.

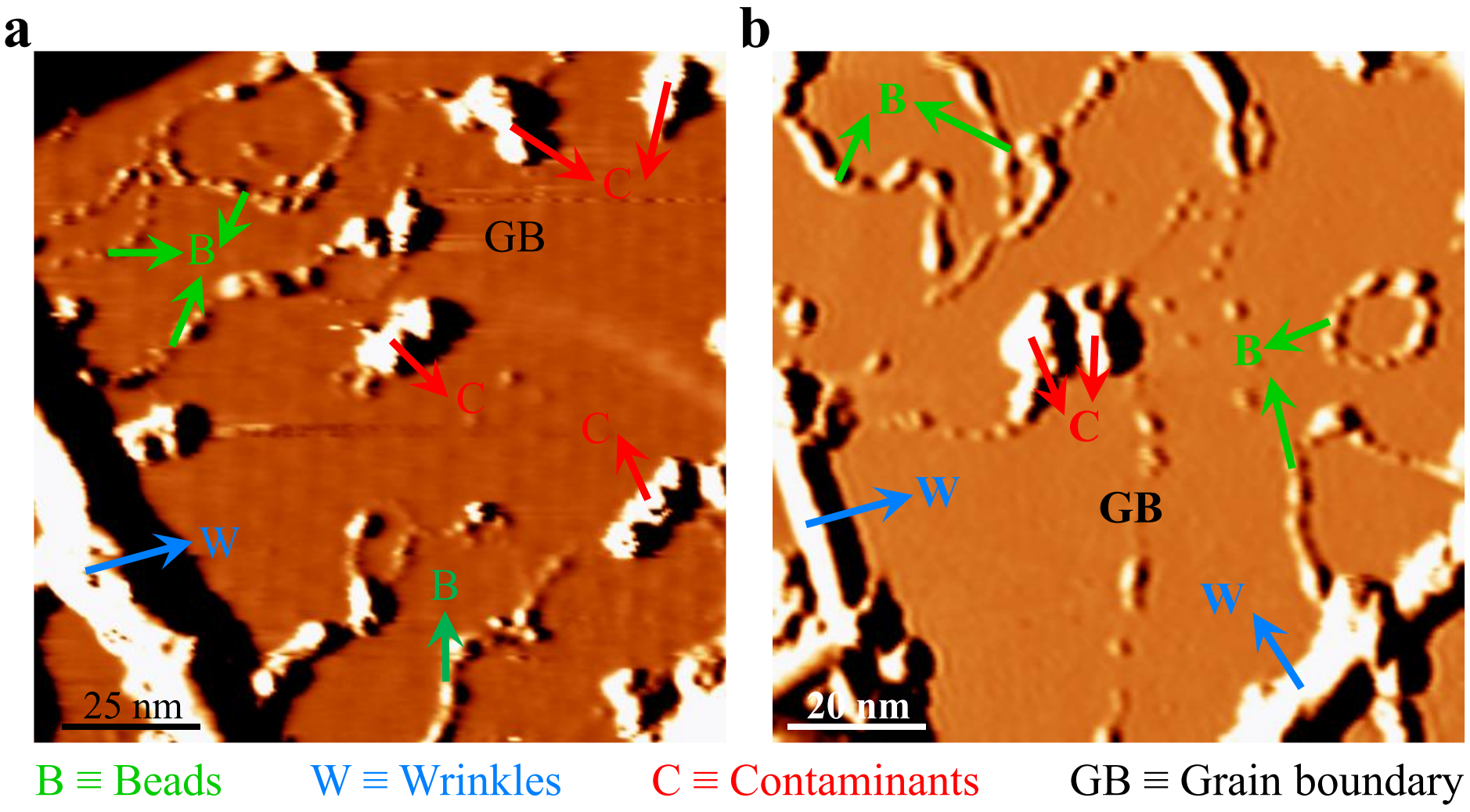

**Figure S20.** General surface morphology of our Gr/SiC(000-1) samples studied in the present work. The main surface imperfections are labelled accordingly. A derivative filter has been applied to all images to enhance the visibility of the features. Acquisition parameters: a) $\Delta f$ = -4 Hz, $V_{CPD}$ = 130 mV. b) $\Delta f$ = -0.8 Hz, $V_{CPD}$ = -40 mV.

**Supplementary note 10: Graphene grain boundaries visualized in topographic ncAFM and STM imaging mode**

In principle, graphene grain boundaries can be characterized using various scanning probe microscopy techniques and their different imaging modes. In this study, we demonstrate that the intrinsic chemical structure of these boundaries has been successfully imaged only through the constant height imaging mode of ncAFM, following tip-apex conditioning as described in Supplementary note 9. Notably, this structure remains inaccessible via constant *frequency shift* (topographic) ncAFM or constant current (topographic) STM imaging modes. This limitation is evident in Figure S21 and Figure S22, where we show the grain boundary as imaged in both the topographic imaging mode of ncAFM and STM.

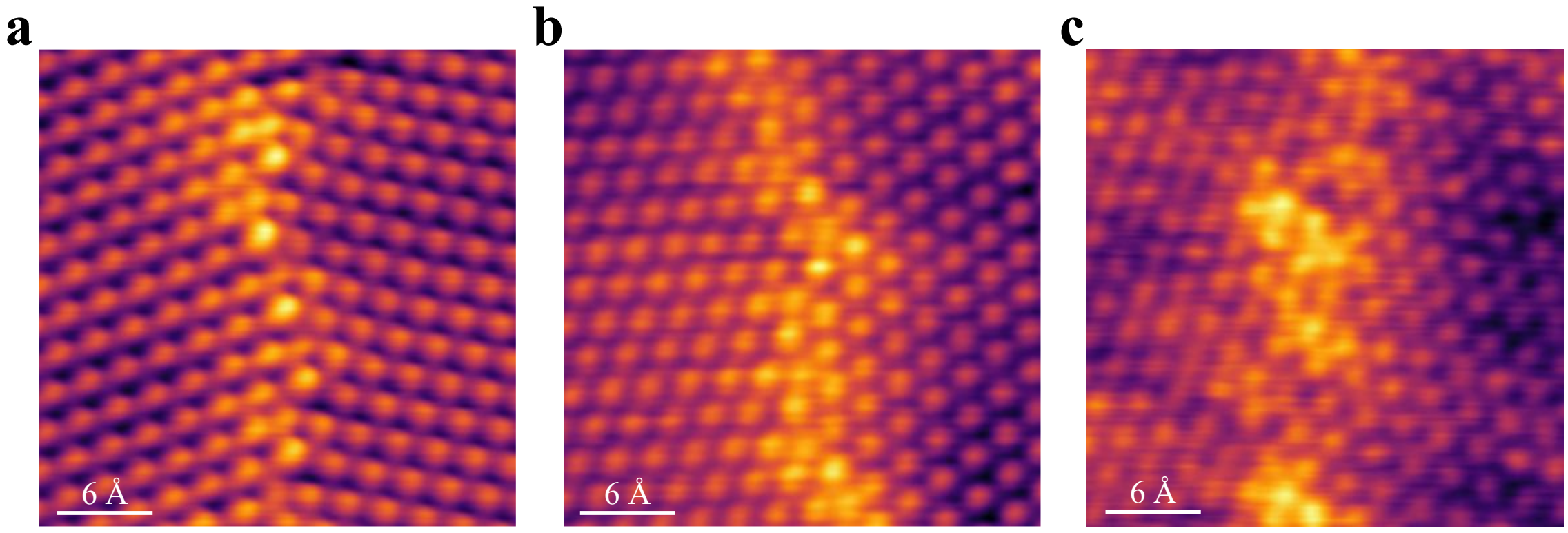

**Figure S21. Graphene grain boundaries visualized in constant $\Delta f$ imaging mode. a-c)** Representative topographic ncAFM images acquired at different GB regions of the sample. As can be observed, the chemical structure of the grain boundaries can not be resolved in this measurement mode. The misorientation angle between the graphene domains is: (a) 31º, (b) 30º, and (c) 35º.

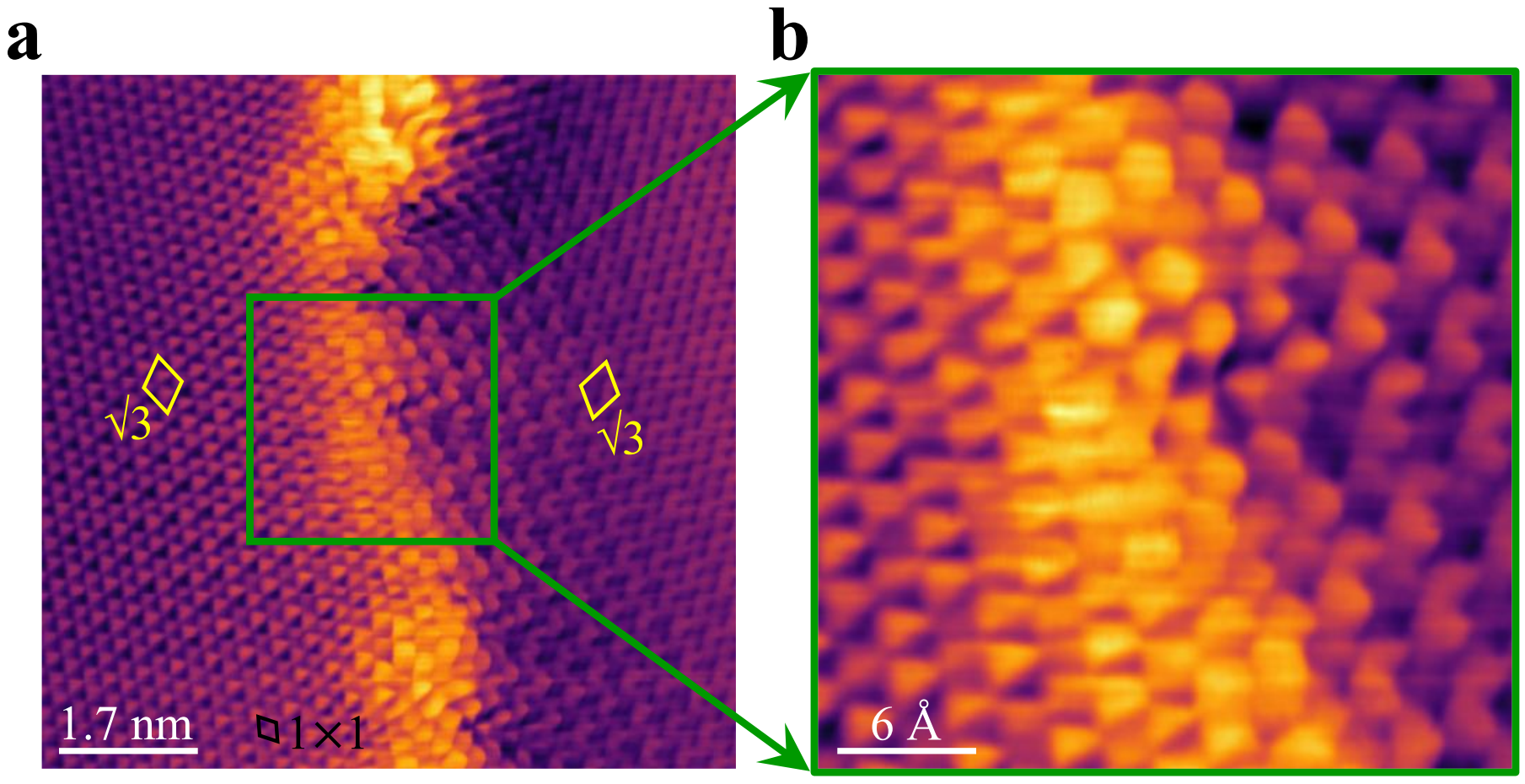

**Figure S22. Graphene grain boundary observed by STM.** a) Constant current STM image of a grain boundary showing intervalley scattering of graphene quasiparticles, evidenced by a (√3×√3)-R30º interference pattern relative to the graphene (1×1) lattice. b) Magnified STM image within the green square region in (a). The chemical structure of GB is unresolved due to STM probes the LDOS, which is dominated by the grain boundary's electronic structure. Acquisition parameters: a,b) $V_s$ = -200 mV, $I_t$ = 0.18 nA. c) $V_{CPD}$ = -40 mV.